# Tutorial: AI-assisted exploration and active design of polymers with high intrinsic thermal conductivity

Xiang Huang[1] and Shenghong Ju[1, 2, a)]

[1] China-UK Low Carbon College, Shanghai Jiao Tong University, Shanghai, 201306, China

[2] Materials Genome Initiative Center, School of Material Science and Engineering, Shanghai Jiao Tong University, Shanghai, 201306, China

[a)] Author to whom correspondence should be addressed: shenghong.ju@sjtu.edu.cn

## ABSTRACT

Designing polymers with high intrinsic thermal conductivity (TC) is critically important for the thermal management of organic electronics and photonics. However, this is a challenging task owing to the diversity of the chemical space and the barriers to advanced synthetic experiments/characterization techniques for polymers. In this Tutorial, the fundamentals and implementation of combining classical molecular dynamics simulation and machine learning (ML) for the development of polymers with high TC are comprehensively introduced. We begin by describing the core components of a universal ML framework, involving polymer datasets, property calculators, feature engineering and informatics algorithms. Then, the process of constructing interpretable regression algorithms for TC prediction is introduced, aiming to extract the underlying relationships between microstructures and TCs for polymers. We also explore the design of sequence-ordered polymers with high TC using lightweight and mainstream active learning algorithms. Lastly, we conclude by addressing the current limitations and suggesting potential avenues for future research on this topic.

# I. INTRODUCTION

Polymers have attracted extensive attention in various fields such as energy,[1] environment,[2] electronics,[3] biologies,[4] medicine[5] and engineering[6] thanks to their light weight, low cost, excellent mechanical ductility, superior biocompatibility, and good chemical and thermal stability.[7-9] However, intrinsic bulk polymers are thermally insulating and have a low thermal conductivity (TC) in a narrow range of 0.1 ~ 0.3 W m$^{-1}$K$^{-1}$,[10-12] which restricts the heat dissipation of organic equipment and severely obstructs the process of miniaturization and integration of flexible electronic and optoelectronic devices.[13-20] Achieving high TC in polymers for industrial applications is an urgent demand, and some progress has been realized recently.[21,22] Different fabrication techniques such as micromechanical stretching,[23-26] electrostatic spinning[27-29] and nanoscale templating[30,31] have been employed in the exploitation of polymers, which effectively improve the ordering and crystallinity of the polymer chains and thus exhibit high TC. Moreover, the construction of thermal networks by engineering interchain interactions such as hydrogen bonding,[32] $\pi$-$\pi$ stacking[33] and side chain modifying[34] in polymer blends and copolymers has also been demonstrated to be beneficial for enhanced TC. Taking polyethylene (PE) as an example, the TC of PE films[25] and nanofibers[24] by mechanical stretching was found to be as high as ~ 62 and ~ 104 W m$^{-1}$K$^{-1}$, respectively, over two or three orders-of-magnitude greater than that of typical polymers.

Despite the engineered polymers that can be produced experimentally to achieve increased TC, this requires a strong chemical background from investigators and is limited by process and characterization instruments. Further, the applicability of different techniques is restricted, e.g. micromechanical stretching is unsuitable for brittle polymers.[35] Accompanied by the evolution of high-performance computers and the revolution of multiscale simulation methodology, in silico experiments are playing an important role in the study of thermal transport in polymers.[36-40] Computational approaches including first-principles calculations[41,42] and molecular dynamics (MD) simulations[43] have led the way in revealing the effect of polymer nanostructures on TC. The first-principles calculation to TC is based on the computation of interatomic force constants via density-functional theory (DFT). On this basis, all relevant phonon properties can be calculated using lattice dynamics and the Boltzmann transport equation.[42] This method has been successfully applied to molecular crystals such as PE,[41] polyvinylidene fluoride[41] and polythiophene,[42] but is challenging to apply in amorphous systems, owing to quantum nuclear motion and their complex primitive cells.

MD simulations employ classical force fields combined with Newtonian mechanics and statistical physics to derive the macroscopic properties of systems.[44] MD simulations can handle polymer systems containing tens of thousands of atoms, which are widely used in polymer thermal transport studies, not only to predict the TC of hierarchical structures, but also to probe the linkages between microstructures and TCs. Using MD simulations, an individual PE chain exhibits a very high TC of 350W m$^{-1}$K$^{-1}$, even divergent in some cases.[45-47] These results are encouraging and inspire researchers to make further efforts to develop polymers with high TC. Moreover, properties of polymers such as molecular weight,[48,49] chain length,[50-54] side chains,[34,55-57] and chain conformation;[49,58-61] intra-chain effects such as bonds,[62-64] angles,[65,66] and dihedrals;[67-69] inter-chain behaviours such as molecular cross-linking,[70-75] hydrogen bonding networks[76-79] and $\pi-\pi$ stacking[56,80-82] on TC were extracted in some separate MD simulations. However, current investigations on the thermal transport mechanisms of polymers have mainly focused on common polymers such as PE, polytetrafluoroethylene (PTFE), polyvinylidene fluoride, and other conjugated structures.[83]

For a long time, researchers have been working on exploring quantitative structure-activity relationships (QSAR) from chemical data, which in turn has enabled the rational design of innovative materials, including polymers, inorganic and biological components.[84] QSAR models reveal empirical, linear or non-linear relationships between descriptors extracted from chemical structures and computational/experimental properties or activities.[85] As modern research methods have facilitated the proliferation in the amount of chemical data, the data-driven research paradigm is of critical importance for QSAR modeling[86-88]. In terms of polymers, the chemical space is enormous,[89] corresponding to potential candidates of small organic molecules as many as 10$^{60}$, whereas the known organic compounds are more than 10$^8$ recorded in the PubChem database.[90-92] Moreover, virtual chemical reactions[93] or generative algorithms[94] for small molecules can create a nearly infinite chemical space. Polymer informatics is a data-centric technology equipped with artificial intelligence and machine learning (ML) as powerful engines to accelerate the optimization of organic materials and the development of novel macromolecules.[95-97] Polymer informatics has been effective in facilitating polymer innovation, and has achieved a series of successful applications, involving optical,[98-100] electrical,[101-104] thermal,[105-108] mechanical[109,110] and other properties. [111-114] Over the past five years, several efforts have been made to apply ML to the exploitation of polymers with high TC (TC > 0.30 W m$^{-1}$K$^{-1}$), with essential contributions for expanding the potential candidates and revealing underlying physical mechanisms.[83,91,115-120]

In this Tutorial, we introduce some development paradigms combining high-throughput MD simulations of and machine ML for high TC polymers in the hope of inspiring new researchers who are interested in becoming involved in this field. We start with a description of three core components in polymer informatics of polymer datasets, TC simulation methods and polymer representations in Section II. Following this, interpretable regression models are constructed for mapping polymer microstructures to TCs in Section III. Next, active optimization algorithms are utilized for the design of polymers with high thermal conductivity in Section IV, including single- and multi-objective cases. Our conclusions and outlooks for this area are provided in Section V.

## II. METHODOLOGY

The principle of polymer informatics is to establish patterns from a sufficient amount of existing or generated polymer data, thus facilitating the design/discovery of new functional polymers with improved target properties.[89] Fig. 1 illustrates a mainstream informatics framework for the development of polymers with high TC, consisting of four elements: 1) Polymer datasets; 2) Polymer modeling and TC calculations; 3) Feature engineering, and 4) Informatics algorithms. In the following, we explain the implementation of the design framework in these four aspects.

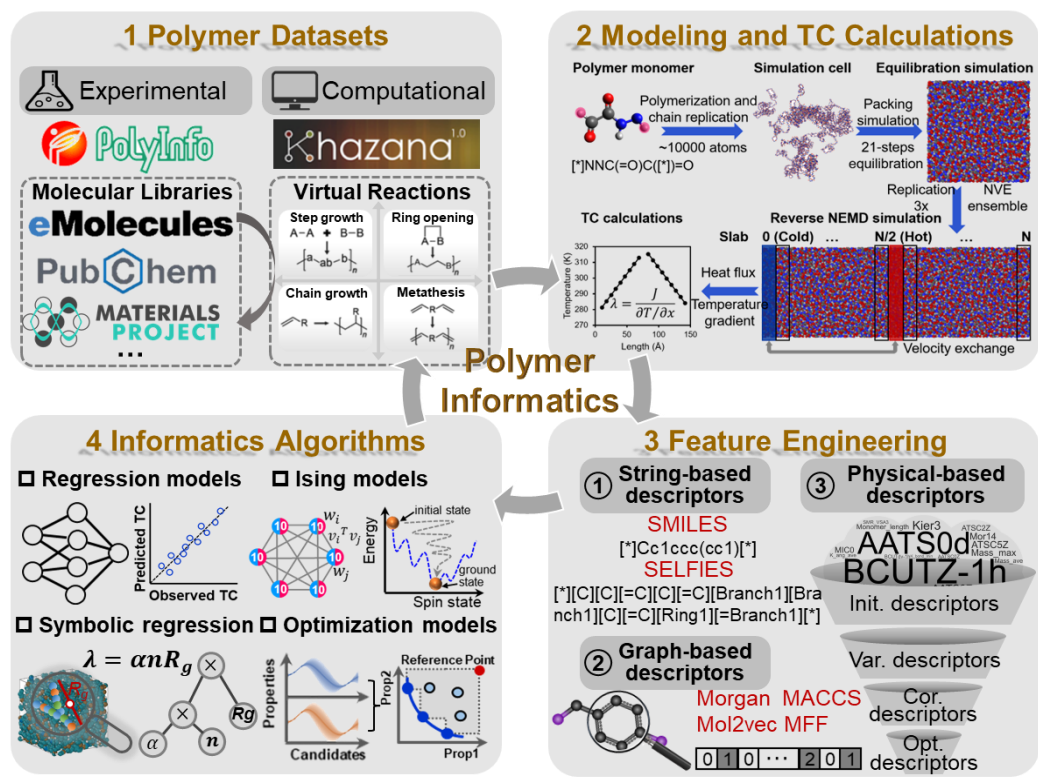

**FIG. 1.** Schematic of machine learning for high thermal conductivity polymer exploitation.

**A. Polymer datasets**

Well-organized and clean data is a fundamental prerequisite for the accomplishment of high-fidelity informatics algorithms. To promote the openness and sharing of scientific data, the findable, accessible, interoperable, reusable (FAIR) principle has been proposed for data storage and management,[121,122] which also greatly contributed to the progression of material genetic engineering.[123] Inorganic materials have a more extensive and accessible database than polymeric materials, such as Materials Project,[124] Atomly,[125] ICSD[126] and AFLOW.[127] These inorganic databases usually follow the FAIR guidelines and provide application programming interfaces (APIs) for access and download. However, the development process of polymer databases is relatively slow and is mostly limited to access. This is attributed to the fact that large polymer primitive cells make experimentation/computation difficult and costly,[6] and empirical nomenclature[128] and diverse forms of expression (strings and graphs) of polymers prevent text-mining techniques from obtaining property data from the published scientific literature.[129]

A few representative databases include PoLyInfo,[130] Khazana,[131] Polymers: A Property Database,[132] Polymer Property Predictor and Database,[133] CAMPUS[134] and PI1M[94] are listed in Table I. PoLyInfo[130] is one of the largest polymer experimental databases, with over 18,000 homopolymers and 7,000 copolymers from about 20,000 scientific literature, including hundreds of properties such as refractive index, dielectric constant and glass transition temperature. Nevertheless, the abundance of different properties is quite variable, for example, more than 8000 homopolymers contain recorded glass transition temperature, while only 84 homopolymers expose TC (accessed 2023.12.14). Moreover, PoLyInfo is contributed by the National Institute for Materials Science of Japan, and is prohibited for downloading large amounts of data. To address this concern, Hayashi et al.[135] selected 1070 amorphous polymers from the PoLyInfo database and calculated the associated 15 properties, including TC, using all-atom classical MD simulations. Ma et al.[94] trained a recurrent neural network based on ~12000 polymers collected from PoLyInfo, and then generated ~1 million polymers to form a benchmark database of PI1M, which covers similar chemical space as the training datasets.

Databases of computational properties of polymers are rarer than experimental databases, of which Khazana[131] is a typical one. The Khazana database was supported by the Ramprasad's group, including DFT computed refractive index, dielectric constant and band gap, and was used as training datasets to build an open informatics platform of Polymer Genome.[136] In Table I, we also describe three databases containing small molecule datasets, namely PubChem,[90] eMolecules[137] and Material

Project.[138] With a grasp of the laws of chemical reactions, virtual reactions can convert small molecules into polymers through ML[139,140] or specific grammar rules.[93,112,141] Yang et al.[112] established a polymer dataset containing more than 8 million hypothetical polyimides formed by known dianhydride and diamine/diisocyanate pairs from PubChem.[90] Kim et al.[93] released a vast polymer database known as the Open Macromolecular Genome, which began with approximately 24 million potential reactant molecules of eMolecules,[137] and then formed synthesizable polymer chemistries compatible with 17 polymerization reactions that cover a variety of step growth chain growth addition, ring opening, and metathesis polymerization reactions.

**TABLE I.** Available organic databases, including mainstream polymer and small molecule datasets.

| No. | Name | Description | Type |
|---|---|---|---|
| 1 | PoLyInfo[130] | PoLyInfo provides various data collected from scientific literature for polymeric material design, including more than 18000 homopolymers and 7000 copolymers. | Polymer datasets |
| 2 | Khazana[131] | Computational materials knowledgebase to store structures and property data created by atomistic simulations | |
| 3 | Polymers: A Property Database[132] | A scientific and commercial information platform for polymers, which contains almost 1000 polymers and 1500 monomers with various properties of mechanical, electrical, thermal, and so on. | |
| 4 | Polymer Property Predictor and Database[133] | Polymer structural and property datasets were collected from the literature through an automated information extraction pipeline. | |
| 5 | CAMPUS[134] | High-quality and comparable material information database with online datasheets for resins from participating material producers. | |
| 6 | PI1M[94] | PI1M contains ~1 million polymers with structural information generated by a recurrent neural network model, which was trained on ~12,000 polymers from the PoLyInfo database. | |
| 7 | PubChem[90] | One of the world's largest open chemical databases, which mostly contains small molecules from hundreds of data sources. | Molecule datasets |
| 8 | eMolecules[137] | An open search-and-fulfillment platform for commercial chemical and biological reagents, which covers over 50 million unique structures and 76 million part numbers. | |
| 9 | Material Project[138] | A well-known material database, it recently integrated more than 170000 molecules studied using density functional theory and can be queried through an OpenAPI-compliant application programming interface. | |

**B. Polymer simulation**

Polymer simulation includes modeling, force field assignment, equilibrium simulation, and TC calculations. While some discrete software such as RDKit,[142] Materials Studio,[143] CHARMM-GUI,[144]

Packmol,[145] Moltemplate,[146] Pysimm[147], Polymer Structure Predictor (PSP),[148] Enhanced Monte Carlo (EMC),[149] and Large-scale Atomic/Molecular Massively Parallel Simulator (LAMMPS)[150] enable the above processes to be realized collaboratively, open-source toolkits that facilitate the building of the entire workflow are of great significance in generating the training data required for polymer informatics. Polymer Molecular Dynamics (PMD) package[151] integrates with PSP, EMC and LAMMPS software to realize polymer modelling and high-throughput MD simulations for various properties, including glass transition temperature, viscosity, TC, and so on. RadonPy[135] is a robust open-source Python library that fully automates the calculation of various properties of polymers using all-atom classical MD simulations. The entire simulation process, including modeling, equilibrium and non-equilibrium MD simulations, can be automated by taking only simplified molecular-input line-entry system (SMILES)[152] string of the polymer repeating unit as input. Furthermore, a computational database has been released containing more than 1000 unique amorphous polymers with various thermophysical properties calculated using RadonPy. Given the powerful and convenient capability of RadonPy, the polymer simulation and ML training data for this Tutorial relies primarily on RadonPy and its derived computational dataset.

**1 *Polymer modeling***

The polymer modeling procedure is illustrated in Fig. 2(a). The SMILES string was given as a unique identifier to distinguish between different polymer structures, where two asterisks denote the connection points, and this was fed as an input parameter to RadonPy. Then, the repeating unit was linked by a self-avoiding random-walk algorithm to form an individual polymer chain.[153,154] The degree of polymerization of the polymer chain is controlled by the total number of atoms, which was uniformly set to around 1000 to ignore the dependence of the physical properties on the molecular weight. After that, the second generation of General AMBER Force Field (GAFF2) force field was assigned to the polymer chain, which is expressed as:[155-159]

$$E = \sum_{\text{bonds}} K_{\text{b}}(r - r_0)^2 + \sum_{\text{angles}} K_{\text{a}}(\theta - \theta_0)^2 + \sum_{\text{dihedrals}} K_{\text{d}}[1 + \cos(n_{\text{d}}\varphi - \delta)] + \sum_{\text{impropers}} K_{\text{i}}(\chi - \chi_0)^2$$
$$+ \sum_{i,j} \frac{q_i q_j}{4\pi\varepsilon_0 r_{\text{ij}}} + \sum_{i,j} 4\varepsilon_{\text{ij}} \left[\left(\frac{\sigma_{\text{ij}}}{r_{\text{ij}}}\right)^{12} - \left(\frac{\sigma_{\text{ij}}}{r_{\text{ij}}}\right)^6\right], \quad (1)$$

where $K_b$, $K_a$, $K_{\text{d}}$ and $K_i$ are the force constants of the bond, bond angle, dihedral angle, and improper angle, respectively; $r$, $\theta$, $\varphi$ and $\chi$ are the bond length, bond angle, dihedral angle, and improper angle, respectively; $r_0$, $\theta_0$ and $\chi_0$ are the equilibration structural parameters of the bond, bond angle, and improper angle, respectively; $n_d$ is the multiplicity and $\delta$ is the phase angle; $q_i$ and $q_j$ are the charges of i-th and j-th atoms; $r_{ij}$ is the distance between atoms *i* and *j*; $\varepsilon_{ij}$ and $\sigma_{ij}$ are the depth of the energy potential and equilibrium distance for Lennard–Jones potential, respectively. In RadonPy, if the bond angle parameters of $K_a$ and $\theta_0$ are missing from the predefined parameter, they were automatically estimated in the same manner as GAFF2.

For obtaining a simulation cell, the single polymer chain was duplicated into 10 copies by translational and rotational operations to prevent overlap with each other, and were placed in a large box with a density of ~0.05 g/cm$^3$. The packing simulation was performed to increase the density of the amorphous systems, to be adjusted to a suitable level. An NVT (constant number of atoms, volume, and temperature) simulation with a Nosé−Hoover thermostat was applied to the system in three sequential stages at a temperature of 300 K, from 300 K to 700 K, and held at 700 K, under periodic boundary conditions (PBC) and a time step of 1 fs. Each of these stages took 1 ns, and all the bonds and angles were constrained by the SHAKE algorithm, resulting in a packaged cell with a density of around 0.80 g/cm$^3$.

Equilibrium simulation was executed for the structural relaxation of amorphous polymers, which follows the 21-step compression/relaxation scheme.[160] During the simulation, by combining NVT and NPT (constant number of atoms, pressure, and temperature) simulations with a Noose-Hoover thermostat, the temperature rise to 600 K and fall to 300 K was repeated for about 1.5 ns while the system was compressed to 50,000 atm and then depressurized to 1 atm. In RadonPy, the amorphous system was considered to be in equilibrium when it satisfies the following conditions: the fluctuations in the total, kinetic, bonding, bond angle, dihedral, van der Waals, and long-range Coulomb energies with relative standard deviations (RSDs) of less than 0.05%, 0.05%, 0.1%, 0.1%, 0.2%, 0.2%, and 0.1%, respectively. At the same time, the RSDs for the fluctuations in density and the radius of gyration were less than 0.1% and 1%, respectively. NPT simulation was run at 300 K and 1 atm with a time step of 1fs. The simulated system was checked for equilibrium states every 50 ns until the equilibrium requirements were realized.[135]

The density $\rho$ of the equilibrated system can be denoted as:

$$\rho = m/\langle V \rangle, \qquad (2)$$

where m is the sum of atomic masses, $\langle V \rangle$ is the time-averaged system volume.

The number density $n$ is calculated using the atoms number $N$ and volume $V$ in the equilibrated system:

$$n = N/V, \tag{3}$$

The radius of gyration $R_g$ is given as:

$$Rg = \sqrt{\frac{1}{\Lambda} \sum_{i=1}^{\Lambda} (r_i - r_m)^2}, \tag{4}$$

where $r_i$ is the position of a repeating unit, $\Lambda$ is the number of repeating units in the polymer chain and $r_m$ is the mean position of these repeating units.

The persistence length $\xi$ can be further obtained as:[51]

$$\xi = \frac{R_g^2 \times 6}{2(2p-1) \times l} + \frac{l}{2}, \tag{5}$$

where $p$ is the degree of polymerization of a polymer chain, and $l$ is approximated as the length of the repeating unit.

## *2 Equilibrium molecular dynamics methods*

Equilibrium molecular dynamics (EMD) simulation is executed in an equilibrium state without temperature gradients, so reasonable structural configurations, careful relaxation and optimization are essential for the accuracy of TC estimation. The TC of polymers in EMD simulation is calculated by the Green-Kubo formalism:[45,161-165]

$$\lambda = \frac{V}{k_B T^2} \int_0^\infty \langle J_x(0) J_x(\tau) \rangle d\tau, \tag{6}$$

where $V$ is the volume of the amorphous system, $k_B$ is the Boltzmann constant, $T$ is the temperature, $J_x$ is the heat flux in the $x$-direction, $\tau$ is the correlation delay time, and $\langle J_x(0) J_x(\tau) \rangle$ denotes the heat autocorrelation function (HACF). Figs. 2(b)-(d) provide a case study of TC calculation for PTFE using EMD. The optimized system contains ~10000 atoms, and we additionally performed the NVE simulation 20 ps for obtaining 10 HACFs with a sampling interval of 2 fs. After the HACFs decayed to 0, the TC of PTFE also stabilized with an average value of 0.27 W m$^{-1}$K$^{-1}$.

## *3 Non-equilibrium molecular dynamics methods*

Non-equilibrium molecular dynamics (NEMD) methods by imposing thermostats (normal NEMD) or swapping the kinetic energy of atoms between two regions (reverse NEMD) in the form of a temperature gradient and the formation of a heat flux. Once the system reaches a steady state, the TC can be derived by Fourier's law:

$$\lambda = -\frac{J}{d_T/d_x}, \quad (7)$$

where $J$ is the heat flux, and $d_T/d_x$ is the temperature gradient in the thermal transport direction.

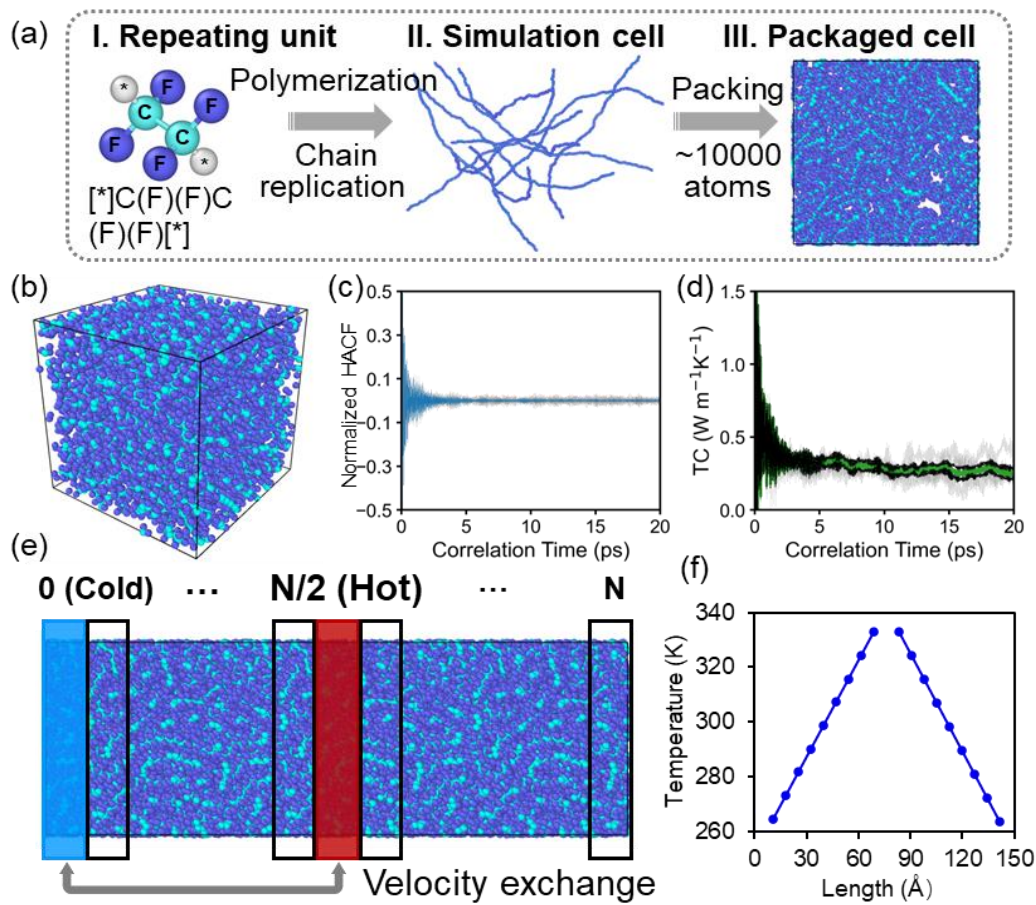

**FIG. 2.** Molecular dynamics simulation for polymer TC calculations. (a) Polymer modeling pipeline. (b) Example snapshot of the relaxed amorphous system. (c) and (d) Heat flux autocorrelation function and TC curves calculated using the Green−Kubo formula. (e) Simulation setup for TC calculation in the non-equilibrium molecular dynamics simulations, where the relaxed system was triple replicated along the heat transport direction, and was then divided equally into *N* slabs, with the red and blue slabs corresponding to the hottest and coldest region. (f) Temperature profile.

In RadonPy, the TC was calculated by reverse NEMD simulation proposed by Müller-Plathe.[166] As shown in Fig. 2(e), the equilibrated system was replicated along the *x*-direction by triplication, and then was divided into *N* (*N*=20) slabs. By exchanging the velocity between the coldest atom in slab *N*/2 and the hottest atom in slab 0, the temperature gradients were formed and recorded for TC evaluation. To prevent temperature shifts after cell replication, the system was initially run at 300K for 2ps. After that, the reverse NEMD simulation was run at 300K for 1 ns with the velocity swapping frequency of 200 fs.[135] According to the exchanged energy *ΔE* obtained using the Müller-Plathe algorithm and the temperature gradient $d_T/d_x$ output by reverse NEMD simulation (Fig. 2(f)), the TC λ can be expressed by:

$$\lambda = \frac{\Delta E}{2 A \Delta t (d_T/d_x)}, \tag{8}$$

where *A* is the cross-sectional area, and $\Delta t$ is the simulation time.

Additionally, the NEMD simulation was implemented for 100 ps for a thermal conductivity decomposition analysis. The energy flux along the direction of unit vectors $J_u$ can be expressed as the contribution of convection (first term) and interatomic interactions (second term):[155,167]

$$J_u = \frac{1}{V} \left\{ \sum_{i \in V} e_i v_{i,u} + \sum_{i \in V} (S_i v_i)_u \right\}, \tag{9}$$

where $v_i$ is the velocity of the atom, $e_i$ is the potential and kinetic energy of each atom, *i* is the index of atoms, and $S_i$ is the stress tensor. For components (a, b), the stress tensor $S_{ab}$ can be detailed as:[135]

$$S_{ab} = \sum_{n=1}^{N_p} r_{i0,a} F_{i,b} + \sum_{n=1}^{N_b} r_{i0,a} F_{i,b} + \sum_{n=1}^{N_a} r_{i0,a} F_{i,b} + \sum_{n=1}^{N_d} r_{i0,a} F_{i,b} + \sum_{n=1}^{N_i} r_{i0,a} F_{i,b} \\ + Kspace(n_{i,a}, F_{i,b}), \tag{10}$$

The stress tensor was divided into six parts of the contributions pairwise, bond, angle, dihedral, improper and K-space, respectively. where $F_i$ is the force acting on atom *i* due to the interaction, $r_{i0}$ stands for the relative position of atom *i* to the geometric center of its interacting atoms, $N_p$, $N_b$, $N_a$, $N_d$, and $N_i$ are the number of interactions of pairs, bonds, bond angles, dihedral angles, and improper angles, respectively. The normalized TC contribution $\eta$ from the equivalent of the heat flux terms is:

$$\eta = J_{par}/J_{tot}, \tag{11}$$

where $J_{tot}$ is the total heat flux calculated by Eq. (9), $J_{par}$ is the partial heat flux solved by Eqs. (10) and (11).

**C. Feature engineering**

Polymer descriptors translate structural and chemical information about polymers into a machine-readable form for ML model training.[168] Successful descriptors are required to uniquely, completely and minimally express polymer information and are a prerequisite and important condition for guaranteeing high accuracy of ML models. To date, many polymer representations have emerged to be exploited for polymer informatics, which can be categorized into three categories[169]: string-based descriptors, graph-based descriptors and physical-based descriptors, as illustrated in Fig. 3.

*1 String-based descriptors*

String-based descriptors are efficient and convenient line notations, such as SMILES[152,170] and self-referencing embedded strings (SELFIES).[171,172] SMILES is a popular polymer representation that uniquely encodes atoms, bonds, rings and branches in polymer monomers by ASCII string. SMILES allows a relatively uniform expression of polymers, does not depend on a strong chemical background, and has good readability for both humans and computers. Thus, SMILES is widely used in data storage,[136] modeling[135,147,148,173] and ML[168,174] of polymers, and has become a standard tool in computational chemistry. However, along with the progress of polymer informatics, SMILES has also exposed some limitations. For instance, in some inverse design tasks based on evolutionary or deep learning algorithms, a lot of the generated SMILES cannot correspond to valid polymers.[175] An improved version of SELFIES[93] was proposed to address this issue, which can represent every polymer and be directly applied in arbitrary ML models without the adaptation of the models. Moreover, SMILES cannot usually be directly fed into the regression models, and need to be further transformed through one-hot encoding[106] or chemical language models.[176-178]

*2 Graph-based descriptors*

Graph-based descriptors are organic chemical representations based on topological information, involving substructure statistics, interatomic connections, and relative positional relationships. Molecular access system (MACCS) keys[179] are one of the most commonly used structural keys, and have been integrated into some open-source cheminformatics software, including RDKit,[142] CDK[180] and OpenBabel.[181] The generated MACCS keys contain 167 bits, the first bit determines whether the molecule has a predefined feature (if exists is set to 1, else is set to 0), and the last 166 bits correspond to 166 substructures. Morgan fingerprints, also known as extended connectivity fingerprints, are adapted from the Morgan algorithm and are one of the most popular fingerprints in chemical information.[182-184] The generation of Morgan fingerprints of a polymer monomer requires three steps:[185] 1) Initialization: Initializes each atom to be encoded as a unique integer identifier. 2) Iteration:

In each iteration, the identifier of each atom is updated to a combination of its own and its neighbours' identifiers. The emerging identifiers are hashed to yield a fixed-length bit vector. Once all atoms have been given new identifiers, the old ones are replaced and the new identifiers are 3) Post-processing: After a prespecified number of iterations, duplicate atom identifiers are removed, and the Morgan fingerprints are formed by the retained identifiers. Mogan fingerprints for polymer representation have the advantages of high efficiency, convenience and absence of a pre-training process, but their feature dimensions are very high and sparse, and possibly introduce bit collisions caused by the hashing process. Mol2vec[186] is an unsupervised method inspired by natural language processing techniques, that considers compound substructures derived from the Morgan algorithm as words and compounds as sentences. Ma at el.[116,187] trained the Mol2vec model on more than 1 million monomer structures from a combination of PoLyInfo and PI1M databases, and obtained a pre-trained model called polymer embedding. Polymer embedding is a 300-dimensional continuous-valued vector, and has been successfully used for the predictions of polymer density, melting temperature, glass transition temperature and TC, respectively. Morgan fingerprint with frequency (MFF)[112] is an expansion of Morgan fingerprints, which captures the frequency of chemical moieties (substructures) present in monomers. Across the entire explored chemical space, substructures (identifiers) with a frequency of occurrences larger than a predefined threshold are preserved to compose descriptor vectors. MFF has much lower feature dimensions compared to Morgan fingerprints, which can effectively suppress the overfitting of ML models and is widely utilized in the prediction of various properties of polymers.[110,111,188]

*3 Physical-based descriptors*

Exploring the collection of physically independent descriptors for characterizing molecular structures is important for qualitative structure-property relationship building and provides a more intuitive guide to molecular performance assessment.[189] Benefiting from the advances in the feature engineering of drug-like molecules, some chemoinformatics software[190-192] is available for the automatic calculation of molecular descriptors. For example, Mordred[192] is a mainstream descriptor-calculation software, that can calculate more than 1800 descriptors, including constructional descriptors, 2D topological descriptors and 3D geometric descriptors.[168] However, polymer monomers are different from small molecules due to the presence of connection site information, which prevents the creation of 3D structures and leads to the inability to obtain some geometric descriptors. To compensate for the lack of 3D information, we incorporated molecular force field (FF) parameters as added descriptors and designed automated physical feature engineering for polymer descriptor optimization in our previous work.[83,120] The initial descriptors are a joint collection, which are calculated by the Mordred software and extracted from the parameters of the polymer structural data file after the FF assignment. Afterwards, feature down-selection technology is employed to acquire the optimized descriptors, which consists of three stages:

1) Evaluate the numerical fluctuations of each physical descriptor itself using a variance indicator and remove features with low variance, since these features have less impact on target properties. This is beneficial in reducing the complexity and improving the performance of ML models. For a specific physical descriptor, the variance can be denoted as:

$$S^2 = \frac{1}{num-1} \sum_{i=1}^{num} (x_i - \bar{x})^2, \tag{12}$$

where $n$ is the number of candidates, $x_i$ is the value of the feature for each candidate, and $\bar{x}$ is the average of the feature values.

2) Filtering and removal of features with poor correlation to target property by four correlation coefficients of the four metrics of Pearson, Spearman, distance, and maximum information coefficients (MIC). For two variables $x$ and $y$, the Pearson coefficient can be solved by:

$$Pea_{x,y} = \frac{\sum_{i=1}^{num}(x_i - \mu_x)(y_i - \mu_y)}{(num-1)S_x S_y}, \tag{13}$$

$$\mu_k = \frac{1}{num} \sum_{i=1}^{num} k, \tag{14}$$

$$S_k = \sqrt{\frac{1}{num-1} \sum_{i=1}^{num} (k_i - \bar{k})^2}, \tag{15}$$

where $\mu_x$ and $\mu_y$ are the mean values of the two variables $x$ and $y$, respectively. $S_x$ and $S_y$ are the standard deviations of the two variables $x$ and $y$, respectively. $k$ can indicate any of the variables $x$ and $y$.

The Spearman coefficient is defined as:

$$r_s = 1 - \frac{6\sum_{i=1}^{num} d_i^2}{num(num^2-1)}, \tag{16}$$

where $d_i$ is the difference in the ranks of the corresponding variables, i.e., the difference in the positions (ranks) of pairs of variables after the two variables have been individually ranked. In addition, the distance correlation coefficient[193] and MIC[194] are used to measure the correlation of possible nonlinear variables.

3) Feature optimization using ML models to better match model performance. Recursive feature elimination (RFE)[195] is a widely employed technique for feature selection, which selects features by removing features with smaller absolute weights while ensuring accuracy in repetitive training of the ML model.[196]

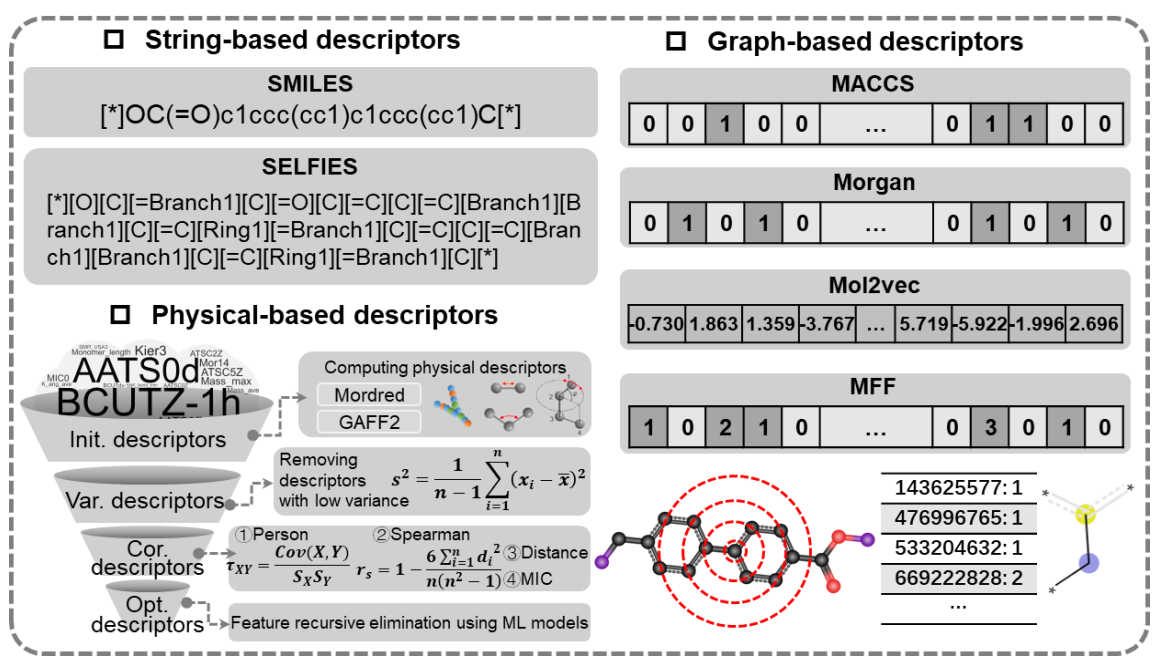

**FIG. 3.** Feature engineering for polymer informatics. It shows the three different options for polymer representation, including string-based, graph-based and physical-based descriptors.

## III. INTERPRETABLE REGRESSION MODELS FOR TC PREDICTION

Establishing the mapping from microstructures to TCs using interpretable ML contributes to the understanding of the intrinsic thermal transport mechanism and guides the design of novel promising structures. Shapley additive explanations (SHAP)[197] is a powerful tool for explaining ML models to alleviate black-box challenges, which links the optimal credit allocation of the model's input features to local explanations. Besides, the symbolic regression (SR)[198-201] technique can construct analytical models of key physical parameters for TC prediction, and intuitively assists in capturing the underlying physical correlations. Some open-source frameworks for SR, such as gplearn,[199] PySR[202] and SISSO,[203] facilitate the discovery of optimal combinations between features and arithmetic symbols, enabling the creation of interpretable explicit mathematical models. Moreover, these tools are user-friendly, and provide helpful step-by-step guidance.[204-206] Here, we presented three case studies of interpretable TC prediction models constructed using the SHAP or SR approaches, respectively.

## A. Regression models trained with physical descriptors

The training of the regression model started with 1051 polymers sourced from a computational database, and all candidates were labelled TCs by NEMD simulations in RadonPy.[135] We then calculated 325 initial descriptors, of which 294 are Mordred-based descriptors and 31 are MD-based descriptors. The down-selection process is displayed in Fig. 4(a). The threshold for variance assessment was 0.10 and a total of 202 descriptors were reserved. A weight assignment mechanism was developed for filtering descriptors, each metric was assigned a factor of 0.25, and descriptors with a cumulative weighting factor of 1 were retained. That is, the descriptor is valid only if all of the four correlation coefficients reach the corresponding thresholds. We filtered out 53 descriptors with a cumulative factor of 1 using the thresholds of 0.050, 0.050, 0.213 and 0.186 for Pearson, Spearman, Distance and maximum information coefficients. Ultimately, the random forest (RF) model combined with RFE was employed in the Scikit-learn[207] for descriptor optimization, where 25 descriptors were determined based on the evaluation of the model's accuracy, as listed in Table II.

The 1051 polymers were represented by optimized physical descriptors and randomly splited by training/test set as 80%/20%. We constructed an RF model using those optimized descriptors, where the hyperparameters for RF were optimized with Bayesian optimization with $R^2$ as a target.[208] The Gaussian regression process and acquisition function with ten random pairs of parameters were selected for initial training, and the ideal parameters for each ML model were determined after 100 optimization iterations.[209] Fig. 4(b) compares the TCs predicted by RF with those calculated by NEMD simulations, with the training and test $R^2$ of 0.88 and 0.78, respectively. Moreover, we conducted 20 evaluations for RF models with optimized descriptors. During each evaluation, the training and test data were randomly sampled from a total of 1051 benchmark data at a ratio of 80/20%. The test $R^2$ of 20 RFs is 0.72±0.02 (mean value and one standard deviation). The prediction error of the RF model in the high TC region (TC > 0.40 W $m^{-1}K^{-1}$) is relatively large, since the percentage of these candidates among 1051 polymers is small (~ 2.28%).[120] In addition, we verified the robust performance of the optimization descriptors and on other ML models such as multilayer perceptron and kernel ridge regression, and confirmed their superiority over graph descriptors of MACCS keys and Morgan fingerprints in our previous work.[120]

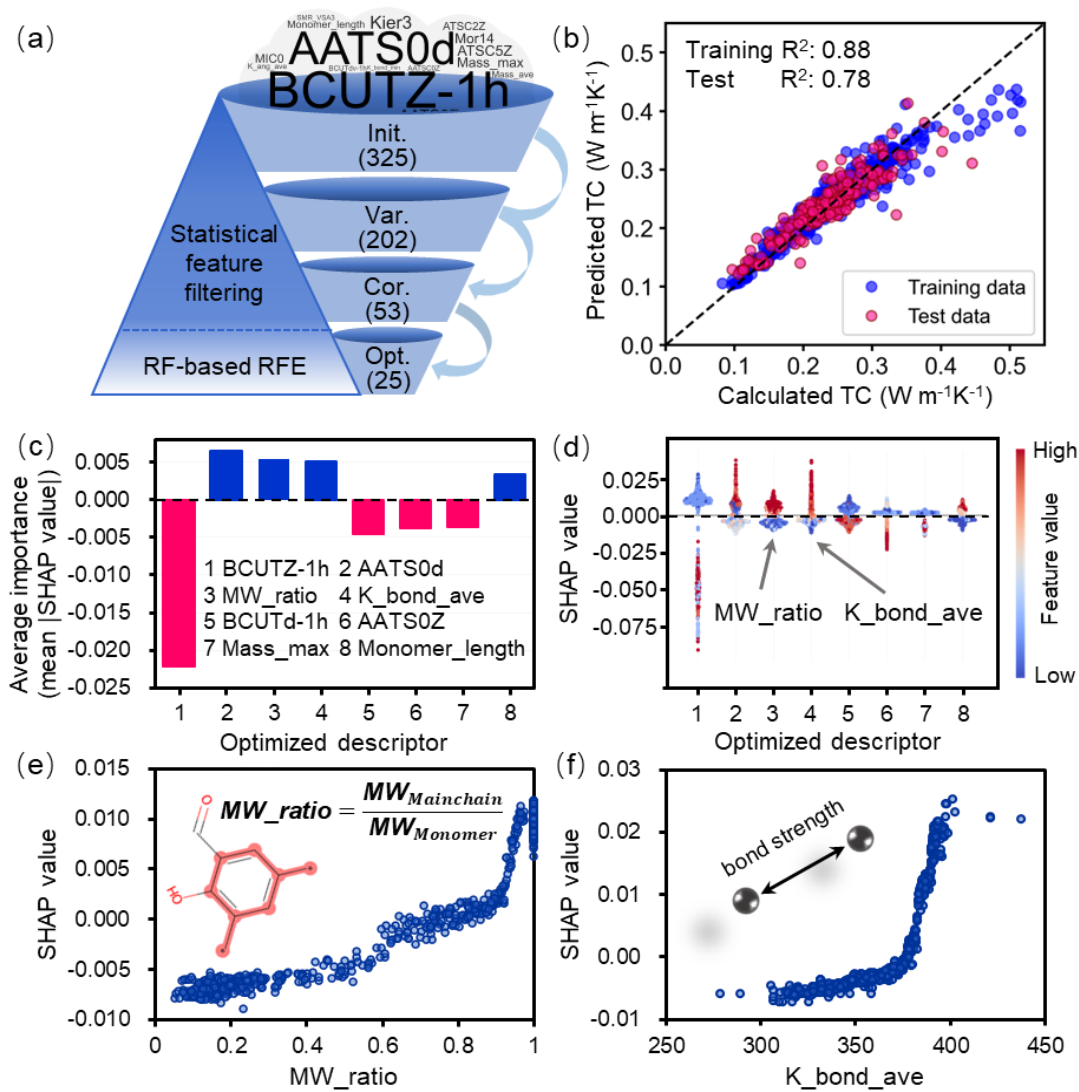

**FIG. 4.** Interpretable machine learning model on physical descriptors and TC. (a) polymer descriptor down-selection process. The initial descriptors (Init.) dimensionality reduction by removing features with low variance (Var.), correlation coefficients filtering (Cor.) and feature recursion elimination to obtain the optimized (Opt.) descriptors. (b) Comparison of MD calculated and RF model predicted TC. (c) Average SHAP importances for optimized descriptors. The blue and red bars indicate positive and negative importance (d) Impact of each optimized descriptor on TC. (e) and (f) SHAP value for the MW_ratio and K_bond_ave of the train set polymer as the functions of descriptor value. The MW_ratio represents the ratio of the molecular weight of the backbone to the total molecular weight of the monomer, and the K_bond_ave indicates the average of different bond force constants.[120]

**TABLE II.** Description of 25 optimized descriptors.[120]

| No. | Labels | Description | Source |
| --- | --- | --- | --- |
| 1 | BCUTZ-1h | First heighest eigenvalue of Burden matrix weighted by atomic number | Mordred |
| 2 | AATS0d | Averaged moreau-broto autocorrelation of lag 0 weighted by sigma electrons | Mordred |
| 3 | MW_ratio | Ratio of mainchain molecular weight to monomer molecular weight | MD |
| 4 | K_bond_ave | Average of different bond force constants in monomer | MD |
| 5 | BCUTd-1h | First heighest eigenvalue of Burden matrix weighted by sigma electrons | Mordred |
| 6 | AATS0Z | Averaged moreau-broto autocorrelation of lag 0 weighted by atomic number | Mordred |
| 7 | Mass_max | Maximum atomic mass in a monomer | MD |
| 8 | Monomer_length | Monomer length after relaxation | MD |
| 9 | Mor02 | 3D-MoRSE (distance = 2) | Mordred |
| 10 | ATSC5Z | Centered moreau-broto autocorrelation of lag 5 weighted by atomic number | Mordred |
| 11 | nHBDon | Number of hydrogen bond donor | Mordred |
| 12 | Mor19 | 3D-MoRSE (distance = 19) | Mordred |
| 13 | Kier3 | Kappa shape index 3 | Mordred |
| 14 | ATSC2Z | Centered moreau-broto autocorrelation of lag 2 Weighted by atomic number | Mordred |
| 15 | Mor14 | 3D-MoRSE (distance = 14) | Mordred |
| 16 | Mass_ave | Average atomic mass in a monomer | MD |
| 17 | AATSC2Z | Averaged and centered moreau-broto autocorrelation of lag 2 weighted by atomic number | Mordred |
| 18 | K_ang_ave | Average of different bond angle force constants in monomer | MD |
| 19 | AATSC0Z | Averaged and centered moreau-broto autocorrelation of lag 0 weighted by atomic number | Mordred |
| 20 | SMR_VSA3 | MOE MR VSA Descriptor 3 ( 1.82 <= x < 2.24) | Mordred |
| 21 | MIC0 | 0-ordered modified information content | Mordred |
| 22 | SMR_VSA1 | MOE MR VSA Descriptor 1 (-inf < x < 1.29) | Mordred |
| 23 | VSA_EState4 | VSA EState Descriptor 4 ( 5.41 <= x < 5.74) | Mordred |
| 24 | MIC1 | 1-ordered modified information content | Mordred |
| 25 | nH | Number of H atoms | Mordred |

The trained RF model was explained by the SHAP approach, and the importance of the features was analyzed based on the output SHAP values. Fig. 4(c) exhibits the eight most important physical descriptors related to properties such as atomic number, atomic mass, bond connection, bond strength, sigma electrons, and monomer length. Combined with the distribution of SHAP values for the first eight physical descriptors of training candidates in Fig. 4(d), the promotion/inhibition of those key descriptors for TC can be recognized in general. The most significant descriptor is BCUTZ-1h, which is the first highest eigenvalue of the Burden matrix weighted by atomic number and is associated with atomic and bonding properties.[210] BCUTZ-1h was observed to have a strong positive correlation with the maximum atomic mass (Mass_max) in the monomer.[120] Typically, the presence of large masses of atoms such as chlorine, bromine and iodine, in the system suppresses lattice vibrations, resulting in small phonon group velocities and low TC. We also analyzed two MD-inspired descriptors of MW_ratio and K_bond_ave in detail, as shown in Figs. 4(e) and (f). The MW_ratio is the ratio of the molecular weight of the backbone to the total molecular weight of the repeating unit, and the K_bond_ave indicates the average of different bond force constants. Both descriptors have a positive effect on TC, as polymers with fewer side chains and stronger chain stiffness are favorable for efficient thermal transport.

**B. Deep neural network trained using MFF**

The utilization of substructures of polymers as descriptors is more intuitive than physical descriptors in revealing structure-TC relationships and facilitates the design of new structures. We give a guide to building a deep neural network (DNN) model using 1144 polymer data with MFF. Among the 1144 polymers, the TCs of most of the structures were collected in a computational database,[135] and the rest were calculated in Radonpy using the same setup.[120] Those polymers were expressed in the form of the SMILES, and converted to MFFs. The MFF has 194 dimensions, mapped to the counts of the 194 most frequent substructures in the entire training dataset. For DNN model training, the hyperparameters were optimized by Keras Tuner[211] Toolkit with Adam optimizer, and mean squared error loss in TensorFlow,[212] using the 1144 polymers with a training/testing ratio of 80%/20%. After optimization of hyperparameters, the trained DNN model has four hidden layers with 416, 256, 244 and 256 nodes, respectively; ReLU activation; and dropout of 0.5.

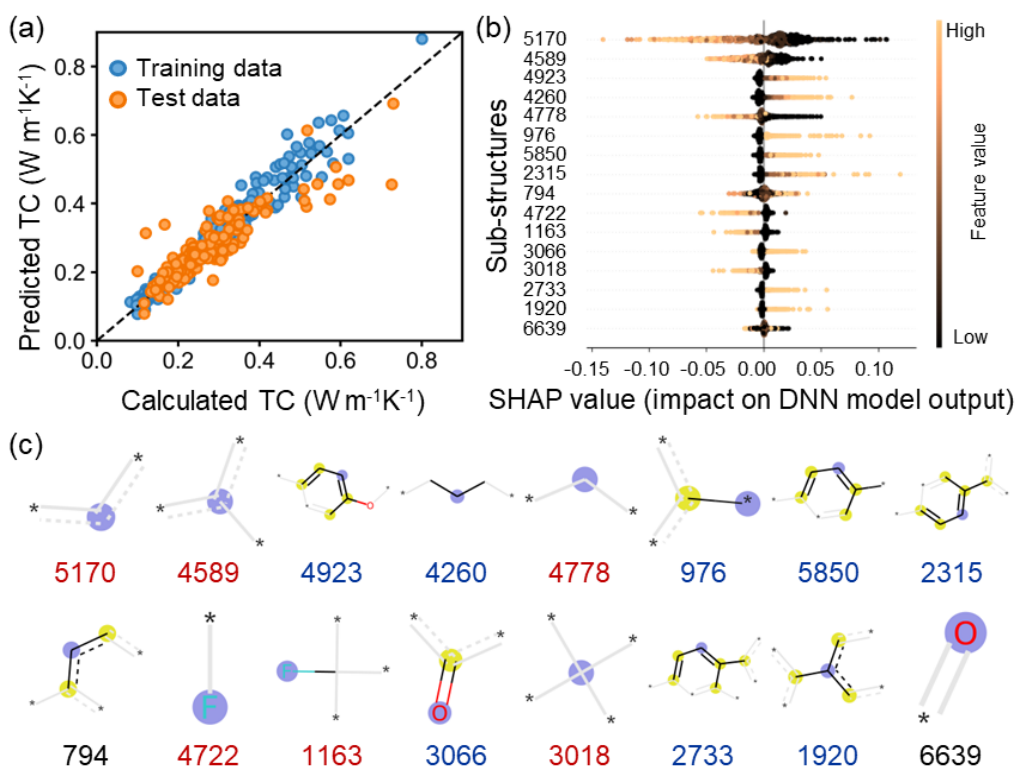

**FIG. 5.** ML model performance and feature importance evaluation. (a) Predicted results for DNN. (b) The interpretations of the DNN model for TC prediction by the SHAP evaluation. (c) The key sub-structures that act on TC, where blue text indicates a positive effect and red indicates an inhibitory effect.[213]

The pairs of DNN-predicted and MD-calculated TCs are plotted in Fig. 5(a), with good consistency and a test root mean square error (RMSE) of 0.040 W m$^{-1}$K$^{-1}$ and test $R^2$ of 0.79. We performed an additional five-fold cross-validation (CV) to evaluate the accuracy of the DNN model. In the five-fold CV, the test $R^2$ of the DNN models is 0.72±0.03 (mean value and one standard deviation)., reflecting that the trained DNN model is reliable.[213] Further, the trained DNN model was interpreted by SHAP, and the role of the most important 16 substructures on TC is shown in Fig.5(b). When the descriptor values follow the same trend as the SHAP values, it suggests that the substructure is contributing to the realization of high TC. Thus, eight substructures were found to have a positive effect on TC, while six substructures suppressed TC, which are marked in Fig.5(c) in blue and red text, respectively. The insights that can be extracted from these substructures coincide with the conclusions previously gained from the RF model trained with physical descriptors in Section III A, namely that conjugated, linear side-chain-free polymers are beneficial for maintaining large chain stiffness and high TC.

Additionally, when the polymer system contains heavy atoms such as fluorine (F), it hinders the efficient heat flux transport, preventing the achievement of high TC.

**C. Symbolic regression for revealing structure-TC correlations**

SR is another interpretable ML method for discovering specific mathematical expressions to match the fit of a dataset.[200] SR simultaneously searches for a set of parameters and the optimal mathematical formula of a function.[214] Reliable training data is critical for SR without requiring massive amounts of data. We calculated the TC of 104 promising amorphous polymers recommended by the RF model,[120] and found that their $R_g$ extracted from equilibrium systems have a strong linear positive correlation with the TCs, as shown in Fig. 6(a). Besides the radius of gyration, we additionally acquired 10 parameters from the equilibrium amorphous systems for SR, as listed in Table III. The SR was implemented in gplearn,[199] and the hyperparameters settings are listed in Table IV. gplearn is a proven tool based on genetic programming that provides a scheme for optimizing mathematical expressions using genetic algorithms. Thus, the main input parameters of the genetic algorithm in gplearn include the optimization generations, the number of formulas produced in each generation, the crossover probability, and the probability of mutation at each node. The formulas were selected based on the criterion of simultaneously having low complexity and high fitting accuracy. The complexity was defined as the length of the formula, with each constant, variable, or operation symbol being represented as a unit of length. Fig. 6(b) statistics of the 3364 mathematical formulae with complexity within 15 and $R^2$ over 0.70 by density plot. The six formulae F1-F6 at the Pareto front are marked by stars, and their corresponding analytic functions are listed in Table V. The Pareto front is the total set of all feasible and Pareto optimal solutions, and is often considered the optimal trade-off between various objectives.[215,216] Overall, formulae with large complexity are more likely to yield large prediction accuracies. Of the six formulas, the smallest complexity is only 5, and the highest precision is 0.876, and their predicted TCs are both in good agreement with those calculated by NEMD, as depicted in Figs. 6(c) and (d). All six formulas capture the positive correlation between $R_g$ and TC, and three of them reflect the fact that it is unfavorable to have large masses of atoms in the systems for TC. $R_g$ was calculated to express the spatial extent of the molecular chains. When the molecular chains in the amorphous system have a high $R_g$, it is beneficial to maintain large rigid chain segments and enhance the heat transport along the chain backbone through intra-chain bonding interactions, thus increasing the TC.[58,120] In addition, it is worth mentioning that since some of the variables utilized for SR, such as $R_g$, were extracted from the equilibrated amorphous systems and are closely related to the TC, resulting in higher prediction accuracy of the obtained analytic mathematical models than that of the RF model in Fig. 4b.

**TABLE III.** Symbol of parameters in symbolic regression.[120]

| No. | Description | Symbol |
|---|---|---|
| $x_0$ | MW_ratio | $M$ |
| $x_1$ | K_bond_ave | $k_{bavg}$ |
| $x_2$ | K_ang_ave | $k_{aavg}$ |
| $x_3$ | Mass_max | $m_{max}$ |
| $x_4$ | nHBDon | $n_H$ |
| $x_5$ | density | $\rho$ |
| $x_6$ | number density | $n$ |
| $x_7$ | radius of gyration | $R_g$ |
| $x_8$ | persistence length | $\xi$ |
| $x_9$ | specific heat capacity at constant pressure | $C_P$ |
| $x_{10}$ | specific heat capacity at constant volume | $C_V$ |

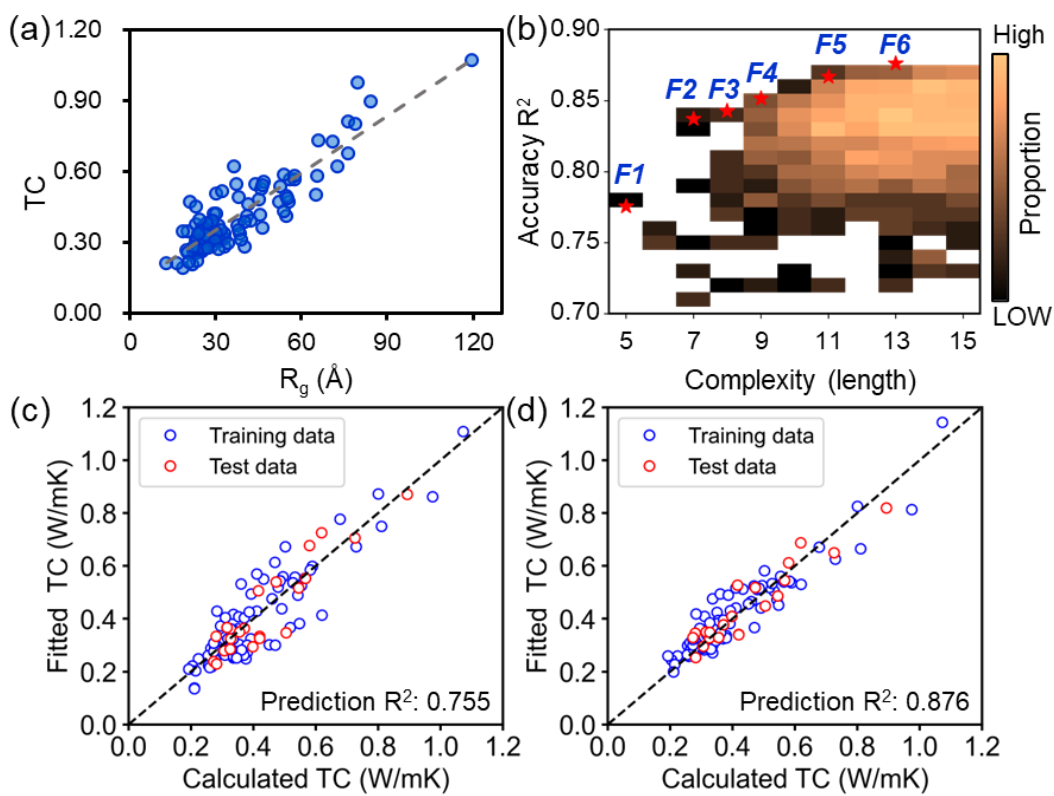

**FIG. 6.** Symbolic regression for analytic model construction. (a) The relationships between the radius of gyration $R_g$ and TC of 104 polymers in this work. (b) Accuracy $R^2$ vs. complexity of 3364 mathematical formulas shown via density plot. The six points of F1-F6 were picked by Pareto search. (c) and (d) Calculated TC vs. fitted TC of formula F1 and F6, respectively.[120]

**TABLE IV.** Hyperparameters setup for symbolic regression.[120]

| Parameter | Value |
|---|---|
| Generations | 300 |
| Population size in every generation | 5000 |
| Probability of crossover (pc) | [0.30,0.90], step=0.05 |
| Probability of subtree mutation (ps) | [(1-pc)/3, (1-pc)/2] (step= 0.01) |
| Probability of hoist mutation (ph) | [(1-pc)/3, (1-pc)/2] (step = 0.01) |
| Probability of point mutation (pp) | 1-pc-ps-ph |
| Function set | $[+, -, \times, \div, \sqrt{x}, \ln x, |x|, -x, 1/x]$ |
| Init_depth | [2, 6], [4, 8], [6, 10], [2, 10] |
| Parsimony coefficient | 0.003, 0.005 |
| Metric | $R^2$ |
| Stopping criteria | 0.900 |
| Random_state | 0, 1, 2, 3, 4 |

**TABLE V.** The six mathematical formulas at the Pareto front in Fig. 6b.[120]

| Number | Formulas | $R^2$ | Complexity |
|---|---|---|---|
| F1 | $0.132 n R_g$ | 0.775 | 5 |
| F2 | $0.01 R_g + 0.598/\xi$ | 0.837 | 7 |
| F3 | $\dfrac{0.043 R_g}{\ln(\xi/0.27)}$ | 0.843 | 8 |
| F4 | $0.085 R_g / (\sqrt{m_{max}\sqrt{\xi}})$ | 0.852 | 9 |
| F5 | $\dfrac{0.081 R_g}{\sqrt{m_{max}\sqrt{\xi} - n_H}}$ | 0.867 | 11 |
| F6 | $\dfrac{R_g}{m_{max}\sqrt{0.024 k_{bavg}\sqrt{\xi - n_H}}}$ | 0.876 | 13 |

The key materials parameters include the K_bond_ave ($k_{bavg}$), Mass_max ($m_{max}$), nHBDon ($n_H$), number density ($n$), radius of gyration ($R_g$) and persistence length ($\xi$).

## IV. ACTIVE DESIGN OF POLYMERS WITH HIGH TC

Active learning is oriented to the design of new polymers driven by target properties, which breaks through the limitations of regression tasks restricted to a fixed exploration chemical space. The inverse design of polymers with high TC can be achieved by some lightweight and smart optimization algorithms, such as genetic algorithm, Bayesian optimization and quantum annealing. In this section, we not only present some cases of polymer design with a single target of high TC, but also additionally consider the synthesizability of polymers in multi-objective optimization trials.

**TABLE VI.** Polymer fragments as basic units for high thermal conductivity polymer design. Each fragment was binary encoded according to serial numbers (No.).[213]

| No. | SMILES of fragments | Code | No. | SMILES of fragments | Code |
|---|---|---|---|---|---|
| 0 | [*]C=C[*] | [00000] | 16 | [*]c1nc2cc3nc([*])[nH]c3cc2[nH]1 | [10000] |
| 1 | [*]CCCCCC[*] | [00001] | 17 | [*]CC(=O)N[*] | [10001] |
| 2 | [*]C#CC=C[*] | [00010] | 18 | [*]CNC(=O)N[*] | [10010] |
| 3 | [*]c1ccc([*])cc1 | [00011] | 19 | [*]C(=O)NNC([*])=O | [10011] |
| 4 | [*]c1ccc([*])[nH]1 | [00100] | 20 | [*]NNC(=O)C([*])=O | [10100] |
| 5 | [*]c1ccc2cc([*])ccc2c1 | [00101] | 21 | [*]c1ccc2oc([*])nc2c1 | [10101] |
| 6 | [*]c1ccc-2c(Cc3cc([*])ccc-23)c1 | [00110] | 22 | [*]c1nc2ccc([*])cc2o1 | [10110] |
| 7 | [*]CO[*] | [00111] | 23 | [*]NC(=O)C=CC(=O)N[*] | [10111] |
| 8 | [*]OC([*])=O | [01000] | 24 | [*]C(=O)C=CC(=O)N-[*] | [11000] |
| 9 | [*]c1ccc([*])o1 | [01001] | 25 | [*]NC(=O)c1ccc([*])cc1 | [11001] |
| 10 | [*]C(=O)C=CC([*])=O | [01010] | 26 | [*]Nc1ccc(C([*])=O)cc1 | [11010] |
| 11 | [*]C(=O)c1ccc(cc1)C([*])=O | [01011] | 27 | [*]N1C(=O)c2ccc([*])cc2C1=O | [11011] |
| 12 | [*]c1cnc([*])nc1 | [01100] | 28 | [*]NC(=O)c1ccc(cc1)C([*])=O | [11100] |
| 13 | [*]Nc1ccc(N[*])cc1 | [01101] | 29 | [*]C(=O)Nc1ccc(NC([*])=O)cc1 | [11101] |
| 14 | [*]c1nc2cc([*])ccc2[nH]1 | [01110] | 30 | [*]n1c(=O)c2cc3c(cc2c1=O)c(=O)n([*])c3=O | [11110] |
| 15 | [*]c1nc2ccc([*])cc2[nH]1 | [01111] | 31 | [*]N1C(=O)c2cc3cc4Cc5cc6cc7C(=O)N([*])C(=O)c7cc6cc5Cc4cc3cc2C1=O | [11111] |

**A. Establishment of benchmark datasets for triblock polymers**

Inspired by the knowledge of the interpretable DNN model outcomes in Section III B and the seen high TC polymers, we constructed a library containing 32 polymer motifs, as listed in Table VI. To ensure the uniqueness of the identification for each fragment, these motifs were binary coded from [00000] to [11111]. Theoretically, we can construct an infinite number of multi-block polymers by controlling the number and order of fragments. However, in consideration of the computational cost and hardware capabilities, we built a benchmark dataset of triblock polymers. Fig. 7(a) depicts the process of producing a triblock polymer, which was characterized by a binary sequence of 15 bits in length. The 15-bit binary sequence was divided equally into three equal parts, each one corresponding to a fragment. The composition of the polymer fragments is directionless, i.e., two polymers consisting of the fragments [0, 2, 29] and [29, 2, 0] are equivalent. The entire benchmark dataset contains 16896 triblock polymers, which were classified into 13 categories referring to the same classification method as PoLyInfo, such as polyolefins, polyethers and polyimides (in Fig.7(c)). The TC of emerging polymers was estimated by a high-fidelity DNN model trained in Section III B and ranged from 0.16 to 1.03 W m$^{-1}$K$^{-1}$, with 42.6% of the TC greater than 0.40 W m$^{-1}$K$^{-1}$ (in Fig.7(d)). Moreover, the synthesizability of these polymers was evaluated by the synthetic accessibility (SA) score.[217] The SA score, which ranges from 1 (easy) to 10 (hard), is calculated by considering both fragment contribution and complexity penalties. This score is used to evaluate the synthesizability of molecules or polymer repeating units. The SA scores for polymers within a range of 2.28 to 6.21, with 6.3% of them having scores less than 3.0. Out of all 16,896 generated polymers, only 4.5% meet the predefined criteria for both TC and SA, which are known as ideal polymers.

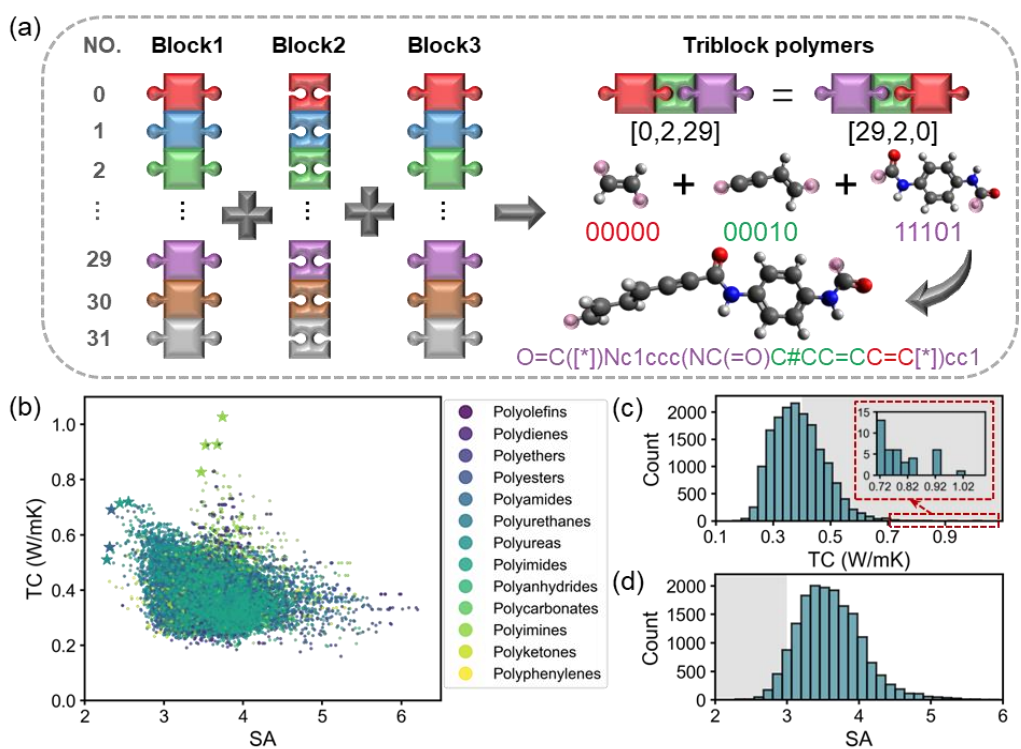

**FIG. 7.** Construction of triblock polymers dataset. (a) Example of the generation of a triblock polymer. (b) SA score versus TC of all 16896 triblock polymers, where stars indicate candidates at the Pareto front. (c) and (d) Distributions of the TC and SA for the whole triblock polymers. The gray strips highlight the statistics of polymers with TC > 0.4 W m$^{-1}$K$^{-1}$ or SA < 3.0.[213]

**B. Single-objective optimization trials**

*1 Genetic algorithm*

The genetic algorithm (GA) is a heuristic search algorithm that is inspired by the biological evolutionary process, and based on the mechanics of natural genetics and selection.[218,219] The operation of the GA algorithm involves various stages of population initialization, fitness evaluation, selection, crossover, and mutation. The core operators of crossover and mutation are illustrated in Fig. 8(a). Once the parents are selected according to the fitness function, the crossover operator combines parents into one or several offspring. After that, the mutation will execute with a predefined probability to increase the diversity in the population.

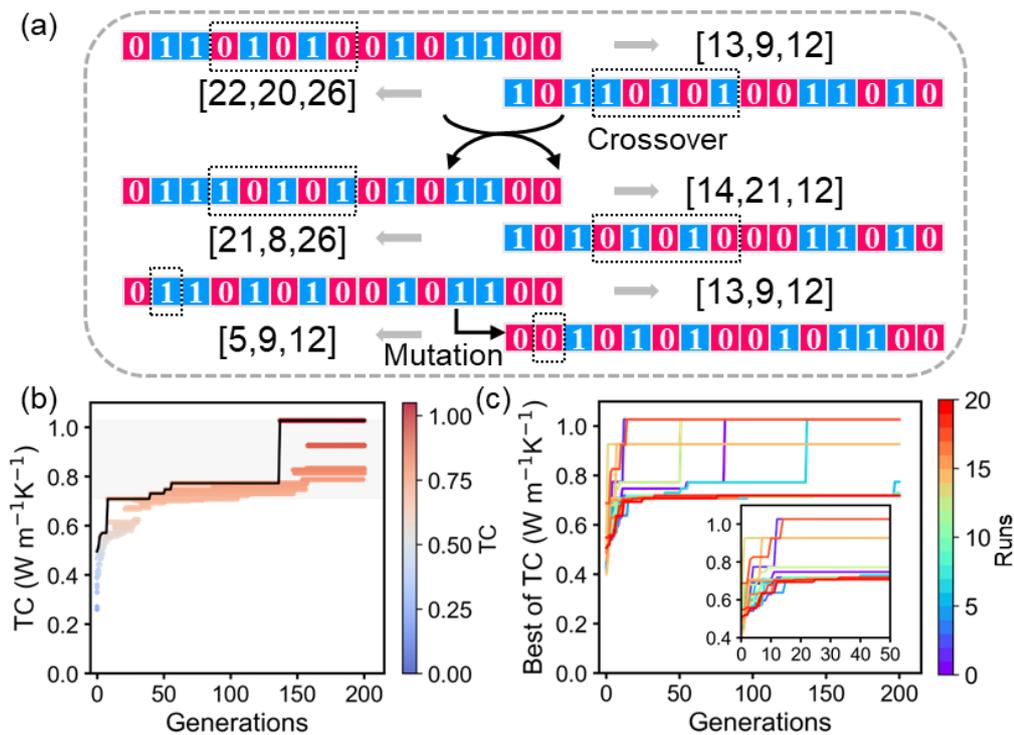

**FIG. 8.** Performance evaluation of genetic algorithm (GA). (a) Illustration of crossover and mutation behaviors in GA. (b) and (c) Convergence of GAs for a single and 20 parallel runs.

In our case, the GA was realized in the pymoo[220] package with simulated binary crossover and polynomial mutation.[221] Fig. 8(b) depicts the convergence curve of the GA in a single optimization run, using the setups of 10 initial random structures, 200 iterations × 10 candidates per batch, and both crossover and mutation probabilities of 100%. Thanks to the inheritance of excellent fragments from the parents, GA can efficiently explore the optimized polymers by only simulating a few candidates (64 non-repeating structures). To probe the effect of the initial structures on the convergence capacity, 20 GAs with different initial structures were executed, and the results are exposed in Fig. 8(c). The TC of the best structures for 20 runs ranges from 0.72 to 1.03 W m$^{-1}$K$^{-1}$, within the top 0.3% of all possible candidates. A total of 6 rounds out of 20 optimizations yielded polymers with the global optimal TC, which indicates that the initial population has a significant influence on the optimization performance of the GA.

*2 Bayesian optimization*

Bayesian optimization (BO) is a method for finding the optimal solution to black-box functions by using sequential design strategies that rely on the probabilistic surrogate models and acquisition functions.[222] Previously, we have released a tutorial on the design of thermal functional materials by coupling thermal transport calculations and BO.[223] Moreover, additional instruction is available to assist in understanding the core components of BO more intuitively.[224] The general process of BO can be described as:[219] 1) Training a surrogate probabilistic model using several structures with the labeled property; 2) Fitting the function using the surrogate model and recommending new structures by the acquisition function; 3) Observing the property of the emerging structures and adding them to the training dataset; 4) Repeating the above process until the predefined iteration times are reached.

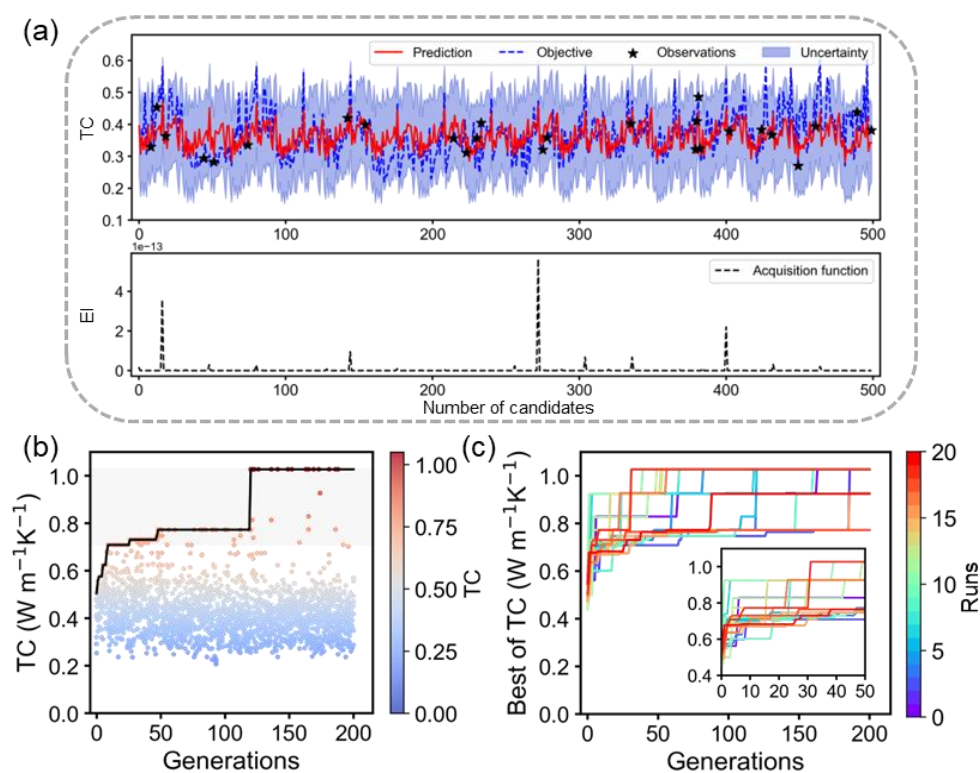

**FIG. 9.** Performance evaluation of Bayesian optimization (BO) algorithm. (a) Gaussian process regression and acquisition functions in BO. (b) and (c) Convergence of BO algorithms for a single and 20 parallel runs.

As shown in Fig. 9(a), the single-task Bayesian optimization with a Gaussian process regression model and the Monte-Carlo acquisition function of qEI was implemented in BoTorch.[225] Fig. 9(b) and (c) count the convergence curves of TCs for single and 20 parallel BO runs, using the hyperparameters set of initial random structures, 200 iterations × 10 candidates per batch, respectively. Along with optimization iterations, BO does search for optimized structures with enhanced TCs. However, many candidates with low TC are invalidly selected due to the uncertainty of Monte Carlo sampling. The best candidates in 20 runs have TC changed from 0.77 to 1.03 W m$^{-1}$K$^{-1}$, of which 14 runs reach the global optimal structure.

*3 Quantum annealing in a quantum virtual machine*

Quantum annealing (QA) is an optimization algorithm assisted by Ising machines to search for the global minimum of a given problem over a given set of candidate solutions (candidate states),[226,227] which are implemented with superconducting qubits, ASICs, GPUs, and so on.[228] QA a has been successfully applied to the design of inorganic and organic materials.[228,229] Ising machine developed specifically for solving quadratic unconstrained binary optimization (QUBO), which is adapted to the binary sequence encoding of polymers. QUBO with *N* bits is described as:[230]

$$\mathcal{H} = \sum_{i=1}^{N}\sum_{j=1}^{N} Q_{ij} q_i q_j, \tag{17}$$

where $q_i$ and $Q_{ij}$ are real-valued parameters, and $Q_{ij} = Q_{ji}$. From Fig.10(a), the factorization machine (FM) was employed as the surrogate model, and the optimal binary solution was identified by trained FM with a quantum annealer. The form of FM is given by:[230]

$$f(\boldsymbol{q}) = \sum_{i=1}^{N} w_i q_i + \sum_{i=1}^{N}\sum_{j=1}^{N}\sum_{k=1}^{K} v_{ik} v_{jk} q_i q_j, \tag{18}$$

where $w_i$ and $v_{ik}$ are real-valued parameters, and the rank K was set to 8 as in the previous literature.[230] Due to hardware limitations, we utilized a sampler called dimod[231] in the Python framework instead of Ising machines to simulate the real quantum annealing. In Figs. 10(c) and (d), although the simulated QA leads to an increase in the TC of the polymers, the TC of optimal polymers below 0.80 W m$^{-1}$K$^{-1}$, and fails to search for the global optimal structure in the 20 parallel optimizations with various initial candidates.

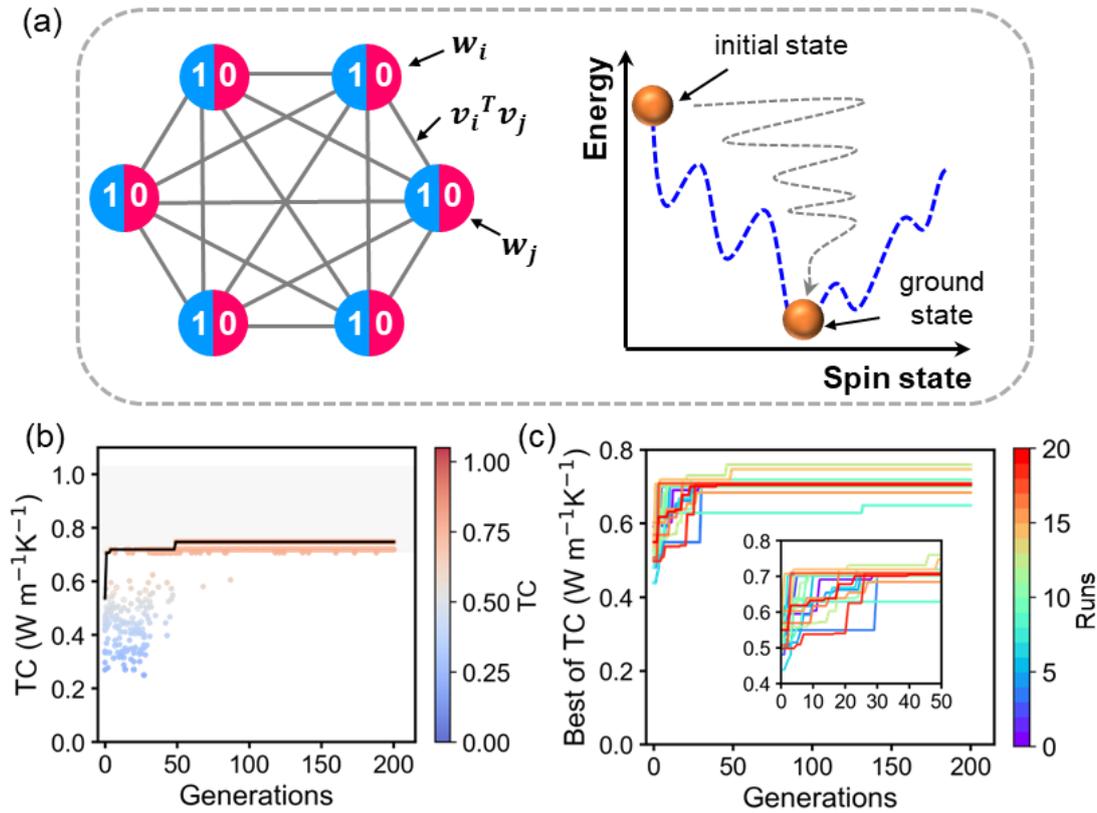

**FIG. 10.** Performance evaluation of quantum annealing (QA) algorithm in a quantum virtual machine. (a) Factorization machine and Ising machine. (b) and (c) Convergence of simulated QA algorithms for a single and 20 parallel runs.

## *4 Performance comparison of three optimization algorithms*

Fig. 11(a) compares the optimization performance of the three algorithms based on the averaged TCs from 20 separate optimization runs at different random states, and the shadows correspond to a standard deviation. Since BO has a robust Gaussian kernel and acquisition function that comprehensively evaluates the TC and uncertainty of the candidates, it has the strongest global optimization capability and the best overall performance. GA is able to inherit some fragments that have a positive effect on TC, thus it enables a rapid increase in the TC of the optimized structures in the early optimization stages, but its ability is confined to the initial populations. Limited to the accuracy of the FM-trained surrogate model and the hardware of Ising machines, the simulated QA performs worse. However, affected by the stochasticity and uncertainty of the Monte Carlo sampler, BO simulated significantly more structures (after de-weighting) in a single optimization run compared to GA and QA (Fig. 11(b)).

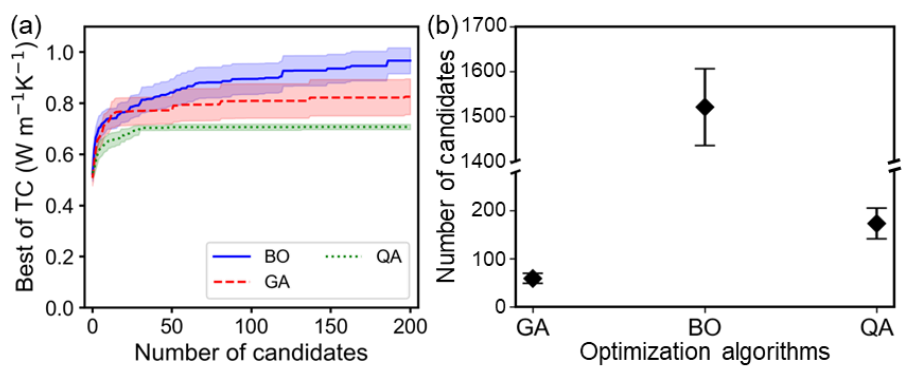

**FIG. 11.** Comparison of different optimization algorithms. (a) Convergence curves of three optimization algorithms, each curve was averaged by the outcomes of 20 runs, and the shading indicates one standard deviation. (b) Number of candidates designed by three optimization algorithms after de-duplication in 20 runs, where the diamonds are the mean values and the error bars represent a standard deviation.

## C. Multi-objective optimization trials

Two state-of-the-art multi-objective optimization algorithms of unified non-dominated sorting genetic algorithm III (U-NSGA-III) and q-noisy expected hypervolume improvement (qNEHVI) were employed for the design of triblock polymers with both high TC and synthetic possibility. As depicted in Fig. 12(a), the U-NSGA-III is a multi-objective evolutionary algorithm (MOEA) that is an updated version of NSGA-III.[232,233] It generalizes different dimensional objective problems by increasing the selection pressure by a scalar selection operator. U-NSGA-III was implemented in the pymoo[220] and kept all hyperparameters with default values. qNEHVI is a multi-objective Bayesian optimization (MOBO) that extends the acquisition function of expected improvement to hypervolume (HV) as an objective. It evaluates samples collected by the Quasi-Monte Carlo (QMC) sampler from the model posterior and identifies the candidate with the largest objective value. qNEHVI was operated in BoTorch,[225] with the base and raw sampling set at 256 and 128 respectively, to enhance computational efficiency.

Figs. 12(b) and (c) illustrate the optimization trajectories for a single run of MOEA and MOBO with 10 random initial structures and 200 iterations × 10 candidates per batch. Nine gray stars indicate the sites of global optimal polymers, while the polymer dots are color-coded according to the generations. The distribution of searched non-duplicated polymer structures in a MOBO run is much denser than those in a MOEA run. qNEHVI integrates HV into the expected improvement acquisition function as an objective to evaluate the randomized QMC samples sourced from the model posterior, generating non-duplicated candidates in almost every generation. This enables the models to break out of local optimal solutions and further increases HV. In contrast, the optimization strategy of U-NSGA-III is inspired by the behavior of genes in organisms that crossover and mutate during evolution. The optimal polymers are designed by randomly selecting parents for matching and introducing a tournament operator. However, the performance of U-NSGA-III is affected by the initial polymer structures, as the optimization process primarily accumulates previous polymer units with positive contributions, making it easy to become trapped in locally optimal solutions.

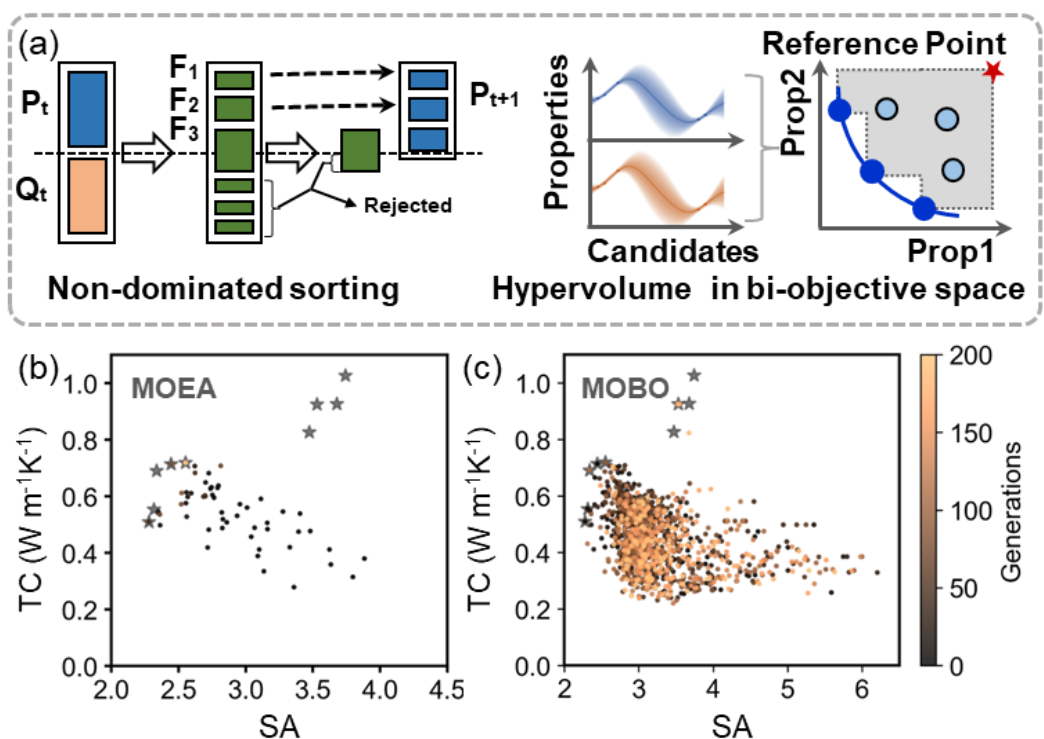

**FIG. 12.** Evaluation of multi-objective optimization algorithms. (a) Core components for U-NSGA-III and qNEHVI. (b) and (c) Optimization trajectories for a single run of MOEA and MOBA with 10 random initial structures and 200 iterations $\times$ 10 candidates per batch.[213]

To obtain statistical outcomes, we performed 20 runs of the MOEA and MOBO algorithms with different initial candidates, respectively. The HV convergence curves are displayed in Figs. 13 (a) and (b). The HVs of U-NSGA-III can rapidly rise to a certain level within 20 generations, but it is difficult to increase again in subsequent generations. However, three qNEHVI runs identified nine global optimal polymers within 200 generations, and almost all of the HVs experienced a secondary boost after reaching a certain level for the first time. This

31 / 48

enhancement difference depends on the stochastic nature of QMC sampling.[215] All the HVs of optimization algorithms reach a referred value calculated by the five ideal global optimal Pareto polymers (TC > 0.4 W m$^{-1}$K$^{-1}$ and SA < 3.0) and the referred point ([0,-10] for TC and SA), although the mean HV of MOBO is greater than that of MOEA (see Figs. 13 (c) and (d)).

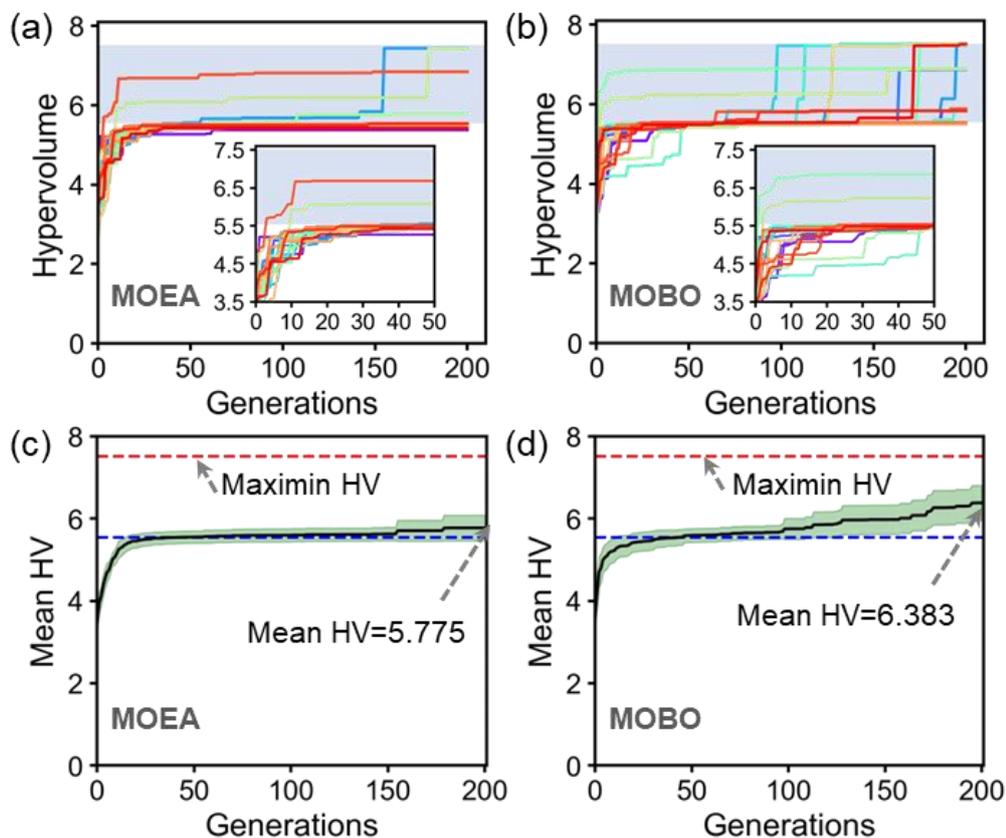

**FIG. 13.** Convergence of multi-objective evolutionary algorithm (MOEA) and multi-objective Bayesian optimization (MOBO) in triblock high TC polymers inverse design. (a) and (b) Convergence curves for 20 runs of MOEA and MOBO. Each optimization run with 10 random initial structures and 200 iterations × 10 candidates per batch. (c) and (d) Mean hypervolume curves for 20 MOEA and MOBO runs. The upper edge of the blue strip or the red dashed line corresponds to the global optimal HV, and the lower edge of the blue strip or the blue dashed line indicates the HV computed from the five ideal global optimal Pareto polymers with the reference points.[213]

Figs. 14(a) and (b) present the number of explored de-duplicate polymers in 20 MOEA and MOBO runs, respectively. The number of unique polymers generated per MOEA run is significantly lower than that of MOBO, with a mean value of approximately 77, which is less than 5.0% of the average value for MOBO. To address this issue, an effective approach is to design high TC polymers through multiple parallel MOEAs with different random states, thereby reducing the influence of initial structures. Furthermore, we calculated the TC of 20 MOEA-designed polymers (red dots) using NEMD in Fig. 14(c), which indeed enhances the Pareto front (marked by stars) formed with 1,144 raw polymers (blue dots). The MD validation of polymers with top 3 TCs are displayed in Fig. 10(d). These polymers are all combinations of fragments of the benzene and 3,5-dihydroimidazo[4,5-f]benzimidazole, with conjugated structures and large backbone stiffness. Consequently, intra-chain interactions of bonds, angles and dihedrals dominate the contribution to TC.

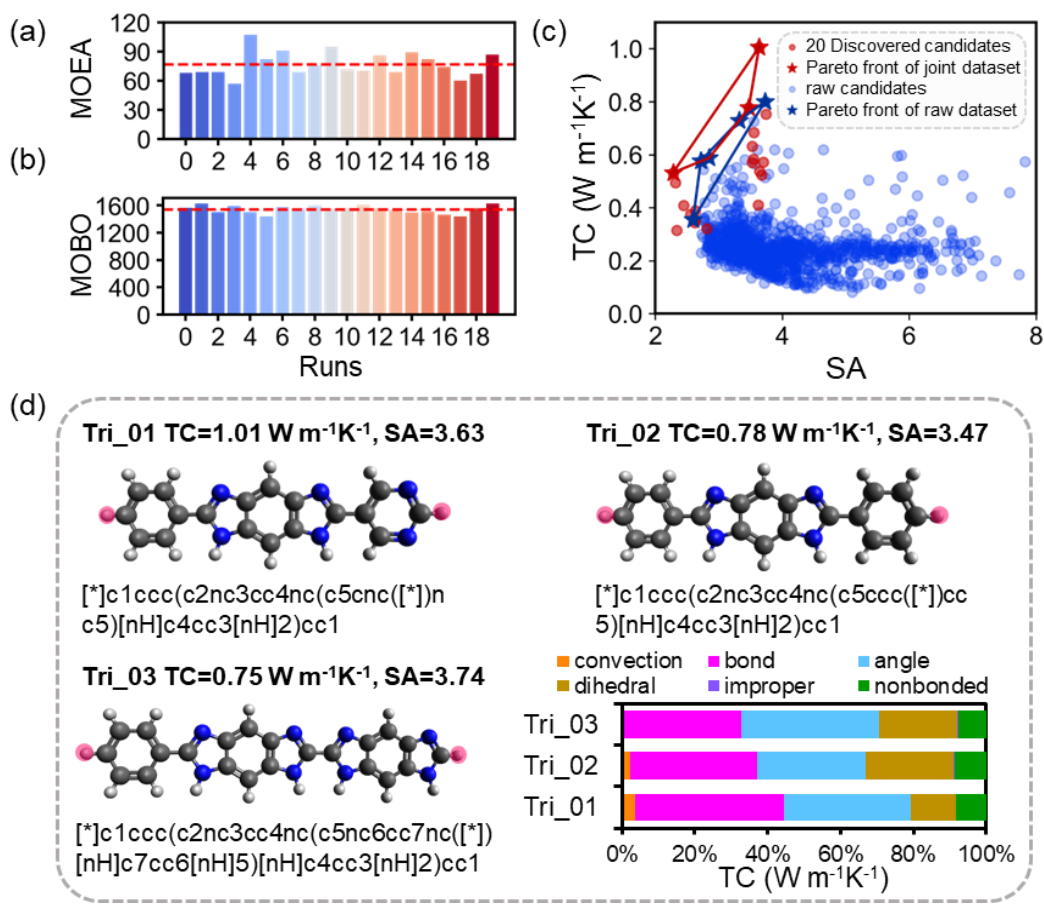

**FIG. 14.** Statistics on the outcomes of inverse design algorithms. (a) and (b) Number of candidates designed by MOEA and MOBO after de-duplication in 20 runs. (c) Pareto front improvement over the 1144 raw training data after adding 20 MOEA-optimized candidates with MD-calculated TC. (d) Quantitative decomposition of TC into contributions from convection and different types of interactions of three high TC polymers.[213]

## V. SUMMARY AND OUTLOOKS

Over the past few years, data-driven informatics algorithms have contributed to a revolution in the materials development paradigm, greatly facilitating the design of polymers and enhancing our understanding of their underlying mechanisms. In this Tutorial, we discuss the basic principles and implementation of ML for the exploitation of high thermal conductivity polymers, covering polymer datasets, polymer modeling and TC calculation, feature engineering, as well as informatics algorithms. We begin by describing the construction of interpretable regression models via physical or graph descriptors, and reveal the mapping between polymer microstructures and TCs. Based on the trained surrogate prediction model and the knowledge derived from the ML, we create a library containing 32 motifs and employ lightweight active learning algorithms to design sequence-ordered triblock polymers with high TCs. We not only focus on designing polymers with a single optimization target of high TC using GA, BO and simulated QA, but also consider the synthetic feasibility of polymers in multi-objective optimization trials that are realized by two state-of-the-art algorithms of U-NSGA-III and qNEHVI, respectively.

Although ML has facilitated the development of macromolecules with high TC, there is still a large gap in satisfying the various demands of realistic engineering applications, which also provides great opportunities for future investigations.

(1) Sufficient high-quality polymer data is a fundamental prerequisite for polymer informatics. Accessible polymer databases are rare compared to inorganic databases, and the recorded data are rather sparse with strict acquisition rules. The development and preservation of publicly accessible polymer databases that adhere to the FAIR principles and encompass a wide range of properties necessitate collaboration and consensus among chemical researchers. In addition, providing APIs for automated batch downloading of data is favored by polymer informatics.

(2) Several open-source software[135,151] enable automated modeling and TC calculations of polymers through classical MD simulation, which promotes the development of polymer informatics. It is crucial to ensure the reliability of the obtained TCs of polymers. Therefore, efforts are being made at the computational level to enhance the accuracy of force fields for MD or to develop first-principles computational methods to be efficiently and economically applicable to the simulation of macromolecular systems.

(3) Most current work on ML in polymer science focuses on the computational TC given the convenience and consistency of the research. Subsequently, with the success of applying ML to automated chemistry experiments,[234-236] we look forward to the emergence of automated platforms that integrate Polymer literature mining, polymer synthesis, TC measurement, data storage and analysis, as well as novel structure generation and TC evaluation. This will enable the expansion of reliable polymer experimental data and the identification of promising polymers with high TCs.

(4) State-of-the-art informatics algorithms are always sought after by the polymer community. Deep learning algorithms such as transfer learning,[91] recurrent neural networks and reinforcement learning[117] have been successfully applied to the exploitation of polymers with high TC. On the one hand, more intelligent molecular generation algorithms are required for the design of polymers with high TC; on the other hand, efforts are made to explore the application of large language models and multitask learning to the design of multifunctional polymers with enhanced TC.

Last but not least, with the rapid advancement of artificial intelligence and automated experiments, we foresee that ML will become a powerful driving force in accelerating the design of advanced polymers to meet the immense demand in various fields. The attractive characteristics of polymers for ML extend beyond TC to encompass other properties such as optical, electrical, and mechanical properties.

## ACKNOWLEDGMENTS

This work was supported by Shanghai Key Fundamental Research Grant (No. 21JC1403300), Shanghai Pujiang Program (No. 20PJ1407500), the National Natural Science Foundation of China (No. 52006134), and the SJTU Global Strategic Partnership Fund (2022 SJTU-Warwick).

## AUTHOR DECLARATIONS

### Conflict of Interest

The authors have no conflicts to disclose.

## DATA AVAILABILITY

Implementation of polymer physical feature engineering and the creation of interpretable machine learning models are available in the GitHub repository: https://github.com/SJTU-MI/APFEforPI, the polymer inverse design cases are available in the GitHub repository: https://github.com/SJTU-

MI/PMBO, and all of the polymer modeling and thermal conductivity calculations achieved in the RadonPy package: https://github.com/RadonPy/RadonPy. More details can be found in our previous publications[120] or on reasonable request from the corresponding author.

## REFERENCES

[1] S. Han, P. Wen, H. Wang, Y. Zhou, Y. Gu, L. Zhang, Y. Shao-Horn, X. Lin, and M. Chen, Nature Materials (2023).

[2] K. L. Law and R. Narayan, Nature Reviews Materials **7,** 104 (2022).

[3] B.-G. Kim, E. J. Jeong, J. W. Chung, S. Seo, B. Koo, and J. Kim, Nature Materials **12,** 659 (2013).

[4] N. Li, Y. Li, Z. Cheng, Y. Liu, Y. Dai, S. Kang, S. Li, N. Shan, S. Wai, A. Ziaja, Y. Wang, J. Strzalka, W. Liu, C. Zhang, X. Gu, J. A. Hubbell, B. Tian, and S. Wang, Science **381,** 686 (2023).

[5] A. Suberi, M. K. Grun, T. Mao, B. Israelow, M. Reschke, J. Grundler, L. Akhtar, T. Lee, K. Shin, A. S. Piotrowski-Daspit, R. J. Homer, A. Iwasaki, H.-W. Suh, and W. M. Saltzman, Science Translational Medicine **15,** eabq0603.

[6] L. Gao, L. Wang, J. Lin, and L. Du, Engineering, https://doi.org/10.1016/j.eng.2023.01.018 (2023).

[7] Y. Guo, Y. Zhou, and Y. Xu, Polymer **233,** 124168 (2021).

[8] X. Xu, J. Zhou, and J. Chen, Advanced Functional Materials **30,** 1904704 (2020).

[9] X. Yang, C. Liang, T. Ma, Y. Guo, J. Kong, J. Gu, M. Chen, and J. Zhu, Advanced Composites and Hybrid Materials **1,** 207 (2018).

[10] S. S. Akhtar, Polymers (Basel) **13** (2021).

[11] P. Zhang, J. Zeng, S. Zhai, Y. Xian, D. Yang, and Q. Li, Macromolecular Materials and Engineering **302,** 1700068 (2017).

[12] B. Liu, Y. Zhou, L. Dong, Q. Lu, and X. Xu, iScience **25,** 105451 (2022).

[13] A. Facchetti, Chemistry of Materials **23,** 733 (2011).

[14] B. Han, B. Liu, G. Wang, Q. Qiu, Z. Wang, Y. Xi, Y. Cui, S. Ma, B. Xu, and H.-Y. Hsu, Advanced Functional Materials **33,** 2300570 (2023).

[15] J. Liao, D. Zhang, and Z. Li, Opto-Electronic Engineering **49,** 210388 (2022).

[16] J. Zhou, R. Li, and T. Luo, npj Computational Materials **9,** 212 (2023).

[17] S. Lin, X. Huang, Z. Bu, L. Yu, T. Dai, Z. Lin, and L. Wang, ECS Journal of Solid State Science and Technology **8,** N93 (2019).

[18] Z. Xu, B. Zhu, X. Liu, T. Lan, Y. Huang, Y. Zhang, and D. Wu, Chemical Engineering Journal **477,** 147246 (2023).

[19] W. Yigen, D. Shuai, L. Xiaojuan, W. Liguo, S. Hongwei, L. Mengjiao, L. Xin, Z. Yang, Z. Guolong, Z. Jianyi, and W. Dezhi, Soft Science **3,** 33 (2023).

[20] Y. Liu, H. Liang, L. Yang, G. Yang, H. Yang, S. Song, Z. Mei, G. Csányi, and B. Cao, Advanced Materials **35,** 2210873 (2023).

[21] X. Wei, Z. Wang, Z. Tian, and T. Luo, Journal of Heat Transfer **143,** 072101 (2021).

[22] X. Wang, W. Wang, C. Yang, D. Han, H. Fan, and J. Zhang, Journal of Applied Physics **130,** 170902 (2021).

[23] M. F. Mina, A. K. M. M. Alam, M. N. K. Chowdhury, S. K. Bhattacharia, and F. J. Baltá Calleja, Polymer-Plastics Technology and Engineering **44,** 523 (2005).

[24] S. Shen, A. Henry, J. Tong, R. Zheng, and G. Chen, Nature Nanotechnology **5,** 251 (2010).

[25] Y. Xu, D. Kraemer, B. Song, Z. Jiang, J. Zhou, J. Loomis, J. Wang, M. Li, H. Ghasemi, X. Huang, X. Li, and G. Chen, Nature Communications **10,** 1771 (2019).

[26] R. Shrestha, P. Li, B. Chatterjee, T. Zheng, X. Wu, Z. Liu, T. Luo, S. Choi, K. Hippalgaonkar, M. P. de Boer, and S. Shen, Nature Communications **9,** 1664 (2018).

[27] W. Kong, Z. Zhang, X. Zhao, and L. Ye, Polymer **290,** 126499 (2024).

[28] J. Ma, Q. Zhang, A. Mayo, Z. Ni, H. Yi, Y. Chen, R. Mu, L. M. Bellan, and D. Li, Nanoscale **7,** 16899 (2015).

[29] C. Lu, S. W. Chiang, H. Du, J. Li, L. Gan, X. Zhang, X. Chu, Y. Yao, B. Li, and F. Kang, Polymer **115,** 52 (2017).

[30] V. Singh, T. L. Bougher, A. Weathers, Y. Cai, K. Bi, M. T. Pettes, S. A. McMenamin, W. Lv, D. P. Resler, T. R. Gattuso, D. H. Altman, K. H. Sandhage, L. Shi, A. Henry, and B. A. Cola, Nature Nanotechnology **9,** 384 (2014).

[31] B.-Y. Cao, Y.-W. Li, J. Kong, H. Chen, Y. Xu, K.-L. Yung, and A. Cai, Polymer **52,** 1711 (2011).

[32] G.-H. Kim, D. Lee, A. Shanker, L. Shao, M. S. Kwon, D. Gidley, J. Kim, and K. P. Pipe, Nature Materials **14,** 295 (2015).

[33] J. Chen, Y. Zhou, X. Huang, C. Yu, D. Han, A. Wang, Y. Zhu, K. Shi, Q. Kang, P. Li, P. Jiang, X. Qian, H. Bao, S. Li, G. Wu, X. Zhu, and Q. Wang, Nature **615,** 62 (2023).

[34] Z. Guo, D. Lee, Y. Liu, F. Sun, A. Sliwinski, H. Gao, P. C. Burns, L. Huang, and T. Luo, Physical

Chemistry Chemical Physics **16,** 7764 (2014).

[35] R. Adhikari and G. H. Michler, Progress in Polymer Science **29,** 949 (2004).

[36] S. Lin, Z. Cai, Y. Wang, L. Zhao, and C. Zhai, npj Computational Materials **5,** 126 (2019).

[37] Y. Jaluria, Applied Thermal Engineering **111,** 1574 (2017).

[38] B. Zhang, P. Mao, Y. Liang, Y. He, W. Liu, and Z. Liu, ES Energy & Environment **5,** 37 (2019).

[39] N. Mehra, L. Mu, T. Ji, X. Yang, J. Kong, J. Gu, and J. Zhu, Applied Materials Today **12,** 92 (2018).

[40] X. Qian, J. Zhou, and G. Chen, Nature Materials **20,** 1188 (2021).

[41] K. Utimula, T. Ichibha, R. Maezono, and K. Hongo, Chemistry of Materials **31,** 4649 (2019).

[42] P. Cheng, N. Shulumba, and A. J. Minnich, Physical Review B **100,** 094306 (2019).

[43] X. Wei, Z. Wang, Z. Tian, and T. Luo, Journal of Heat Transfer **143** (2021).

[44] S. Chmiela, H. E. Sauceda, K.-R. Müller, and A. Tkatchenko, Nature Communications **9,** 3887 (2018).

[45] A. Henry and G. Chen, Physical Review Letters **101,** 235502 (2008).

[46] G. Chen, Nature Reviews Physics **3,** 555 (2021).

[47] W. Lv, R. M. Winters, F. DeAngelis, G. Weinberg, and A. Henry, The Journal of Physical Chemistry A **121,** 5586 (2017).

[48] A. Kiessling, D. N. Simavilla, G. G. Vogiatzis, and D. C. Venerus, Polymer **228,** 123881 (2021).

[49] T. Feng, J. He, A. Rai, D. Hun, J. Liu, and S. S. Shrestha, Physical Review Applied **14,** 044023 (2020).

[50] J. Liu and R. Yang, Physical Review B **86,** 104307 (2012).

[51] X. Wei and T. Luo, Physical Chemistry Chemical Physics **21,** 15523 (2019).

[52] A. Crnjar, C. Melis, and L. Colombo, Physical Review Materials **2,** 015603 (2018).

[53] J. Zhao, J.-W. Jiang, N. Wei, Y. Zhang, and T. Rabczuk, Journal of Applied Physics **113,** 184304 (2013).

[54] A. Chen, Y. Wu, S. Zhou, W. Xu, W. Jiang, Y. Lv, W. Guo, K. Chi, Q. Sun, T. Fu, T. Xie, Y. Zhu, and X.-g. Liang, Materials Advances **1,** 1996 (2020).

[55] D. Luo, C. Huang, and Z. Huang, Journal of Heat Transfer **140** (2017).

[56] X. Wei and T. Luo, Physical Chemistry Chemical Physics **24,** 10272 (2022).

[57] H. Ma and Z. Tian, Applied Physics Letters **110,** 091903 (2017).

[58] X. Wei, T. Zhang, and T. Luo, Physical Chemistry Chemical Physics **18,** 32146 (2016).

[59] R. Muthaiah and J. Garg, Journal of Applied Physics **124,** 105102 (2018).

[60] R. Ma, D. Huang, T. Zhang, and T. Luo, Chemical Physics Letters **704,** 49 (2018).

[61] H. Ma and Z. Tian, Applied Physics Letters **107,** 073111 (2015).

[62] A. B. Robbins and A. J. Minnich, Applied Physics Letters **107,** 201908 (2015).

[63] X. Xie, K. Yang, D. Li, T.-H. Tsai, J. Shin, P. V. Braun, and D. G. Cahill, Physical Review B **95,** 035406 (2017).

[64] E. Lussetti, T. Terao, and F. Müller-Plathe, The Journal of Physical Chemistry B **111,** 11516 (2007).

[65] H. Subramanyan, W. Zhang, J. He, K. Kim, X. Li, and J. Liu, Journal of Applied Physics **125,** 095104 (2019).

[66] H. Ma and Z. Tian, Journal of Materials Research **34,** 126 (2019).

[67] T. Zhang, X. Wu, and T. Luo, The Journal of Physical Chemistry C **118,** 21148 (2014).

[68] T. Zhang and T. Luo, The Journal of Physical Chemistry B **120,** 803 (2016).

[69] T. Luo, K. Esfarjani, J. Shiomi, A. Henry, and G. Chen, Journal of Applied Physics **109,** 074321 (2011).

[70] G. Kikugawa, T. G. Desai, P. Keblinski, and T. Ohara, Journal of Applied Physics **114,** 034302 (2013).

[71] X. Xiong, M. Yang, C. Liu, X. Li, and D. Tang, Journal of Applied Physics **122,** 035104 (2017).

[72] X. Wan, B. Demir, M. An, T. R. Walsh, and N. Yang, International Journal of Heat and Mass Transfer **180,** 121821 (2021).

[73] X. Liu and Z. Rao, Computational Materials Science **172,** 109298 (2020).

[74] M. K. Maurya, J. Wu, M. K. Singh, and D. Mukherji, ACS Macro Letters **11,** 925 (2022).

[75] Z. Zhang and B. Cao, Science China Physics, Mechanics & Astronomy **65,** 117003 (2022).

[76] L. Zhang, M. Ruesch, X. Zhang, Z. Bai, and L. Liu, RSC Advances **5,** 87981 (2015).

[77] H. Zheng, K. Wu, Y. Zhan, K. Wang, and J. Shi, Journal of Polymer Science **61,** 1622 (2023).

[78] N. Mehra, L. Mu, and J. Zhu, Composites Science and Technology **148,** 97 (2017).

[79] W. Li, J. Ma, S. Wu, J. Zhang, and J. Cheng, Polymer Testing **101,** 107275 (2021).

[80] H. Zheng, K. Wu, W. Chen, B. Nan, Z. Qu, and M. Lu, Macromolecular Chemistry and Physics **222,** 2000418 (2021).

[81] W. Shi, Z. Shuai, and D. Wang, Advanced Functional Materials **27,** 1702847 (2017).

[82] M. Sangkhawasi, T. Remsungnen, A. S. Vangnai, R. P. Poo-arporn, and T. Rungrotmongkol, in

*Polymers*; *Vol. 14* (2022).

[83] X. Huang, S. Ma, C. Y. Zhao, H. Wang, and S. Ju, npj Computational Materials **9,** 191 (2023).

[84] E. N. Muratov, J. Bajorath, R. P. Sheridan, I. V. Tetko, D. Filimonov, V. Poroikov, T. I. Oprea, I. I. Baskin, A. Varnek, A. Roitberg, O. Isayev, S. Curtalolo, D. Fourches, Y. Cohen, A. Aspuru-Guzik, D. A. Winkler, D. Agrafiotis, A. Cherkasov, and A. Tropsha, Chemical Society Reviews **49,** 3525 (2020).

[85] H. R. Allcock, Science **255,** 1106 (1992).

[86] A. Agrawal and A. Choudhary, APL Materials **4,** 053208 (2016).

[87] S. Hellberg, M. Sjoestroem, B. Skagerberg, and S. Wold, Journal of Medicinal Chemistry **30,** 1126 (1987).

[88] T. I. Oprea and J. Gottfries, Journal of Combinatorial Chemistry **3,** 157 (2001).

[89] L. Chen, G. Pilania, R. Batra, T. D. Huan, C. Kim, C. Kuenneth, and R. Ramprasad, Materials Science and Engineering: R: Reports **144,** 100595 (2021).

[90] S. Kim, P. A. Thiessen, E. E. Bolton, J. Chen, G. Fu, A. Gindulyte, L. Han, J. He, S. He, B. A. Shoemaker, J. Wang, B. Yu, J. Zhang, and S. H. Bryant, Nucleic Acids Research **44,** D1202 (2015).

[91] S. Wu, Y. Kondo, M.-a. Kakimoto, B. Yang, H. Yamada, I. Kuwajima, G. Lambard, K. Hongo, Y. Xu, J. Shiomi, C. Schick, J. Morikawa, and R. Yoshida, npj Computational Materials **5,** 66 (2019).

[92] C. M. Dobson, Nature **432,** 824 (2004).

[93] S. Kim, C. M. Schroeder, and N. E. Jackson, ACS Polymers Au **3,** 318 (2023).

[94] R. Ma and T. Luo, Journal of Chemical Information and Modeling **60,** 4684 (2020).

[95] W. Sha, Y. Li, S. Tang, J. Tian, Y. Zhao, Y. Guo, W. Zhang, X. Zhang, S. Lu, Y.-C. Cao, and S. Cheng, InfoMat **3,** 353 (2021).

[96] K. Hatakeyama-Sato, Polymer Journal **55,** 117 (2023).

[97] K. Sattari, Y. Xie, and J. Lin, Soft Matter **17,** 7607 (2021).

[98] J. P. Lightstone, L. Chen, C. Kim, R. Batra, and R. Ramprasad, Journal of Applied Physics **127,** 215105 (2020).

[99] A. Mishra, P. Rajak, A. Irie, S. Fukushima, R. K. Kalia, A. Nakano, K.-i. Nomura, F. Shimojo, and P. Vashishta, Applied Physics Letters **123,** 121901 (2023).

[100] M. A. F. Afzal, C. Cheng, and J. Hachmann, The Journal of Chemical Physics **148,** 241712 (2018).

[101] Y. Wang, T. Xie, A. France-Lanord, A. Berkley, J. A. Johnson, Y. Shao-Horn, and J. C. Grossman,

Chemistry of Materials **32,** 4144 (2020).

[102] A. Mannodi-Kanakkithodi, G. Pilania, T. D. Huan, T. Lookman, and R. Ramprasad, Scientific Reports **6,** 20952 (2016).

[103] L. Chen, C. Kim, R. Batra, J. P. Lightstone, C. Wu, Z. Li, A. A. Deshmukh, Y. Wang, H. D. Tran, P. Vashishta, G. A. Sotzing, Y. Cao, and R. Ramprasad, npj Computational Materials **6,** 61 (2020).

[104] B. K. Wheatle, E. F. Fuentes, N. A. Lynd, and V. Ganesan, Macromolecules **53,** 9449 (2020).

[105] R. Bhowmik, S. Sihn, R. Pachter, and J. P. Vernon, Polymer **220,** 123558 (2021).

[106] L. Tao, V. Varshney, and Y. Li, Journal of Chemical Information and Modeling **61,** 5395 (2021).

[107] A. Alesadi, Z. Cao, Z. Li, S. Zhang, H. Zhao, X. Gu, and W. Xia, Cell Reports Physical Science **3,** 100911 (2022).

[108] K. K. Bejagam, J. Lalonde, C. N. Iverson, B. L. Marrone, and G. Pilania, The Journal of Physical Chemistry B **126,** 934 (2022).

[109] T. Yue, J. He, L. Tao, and Y. Li, Journal of Chemical Theory and Computation (2023).

[110] L. Tao, J. He, N. E. Munyaneza, V. Varshney, W. Chen, G. Liu, and Y. Li, Chemical Engineering Journal **465,** 142949 (2023).

[111] L. Tao, J. He, T. Arbaugh, J. R. McCutcheon, and Y. Li, Journal of Membrane Science **665,** 121131 (2023).

[112] J. Yang, L. Tao, J. He, J. R. McCutcheon, and Y. Li, Science Advances **8,** eabn9545 (2022).

[113] S. M. McDonald, E. K. Augustine, Q. Lanners, C. Rudin, L. Catherine Brinson, and M. L. Becker, Nature Communications **14,** 4838 (2023).

[114] A. J. Gormley and M. A. Webb, Nature Reviews Materials **6,** 642 (2021).

[115] M.-X. Zhu, H.-G. Song, Q.-C. Yu, J.-M. Chen, and H.-Y. Zhang, International Journal of Heat and Mass Transfer **162,** 120381 (2020).

[116] R. Ma, H. Zhang, J. Xu, L. Sun, Y. Hayashi, R. Yoshida, J. Shiomi, J.-x. Wang, and T. Luo, Materials Today Physics **28,** 100850 (2022).

[117] R. Ma, H. Zhang, and T. Luo, ACS Applied Materials & Interfaces **14,** 15587 (2022).

[118] A. Nagoya, N. Kikkawa, N. Ohba, T. Baba, S. Kajita, K. Yanai, and T. Takeno, Macromolecules **55,** 3384 (2022).

[119] T. Zhou, Z. Wu, H. K. Chilukoti, and F. Müller-Plathe, Journal of Chemical Theory and Computation **17,** 3772 (2021).

[120] X. Huang, S. Ma, Y. Wu, C. Wan, C. Y. Zhao, H. Wang, and S. Ju, Journal of Materials Chemistry A **11,** 20539 (2023).

[121] M. D. Wilkinson, M. Dumontier, I. J. Aalbersberg, G. Appleton, M. Axton, A. Baak, N. Blomberg, J.-W. Boiten, L. B. da Silva Santos, P. E. Bourne, J. Bouwman, A. J. Brookes, T. Clark, M. Crosas, I. Dillo, O. Dumon, S. Edmunds, C. T. Evelo, R. Finkers, A. Gonzalez-Beltran, A. J. G. Gray, P. Groth, C. Goble, J. S. Grethe, J. Heringa, P. A. C. 't Hoen, R. Hooft, T. Kuhn, R. Kok, J. Kok, S. J. Lusher, M. E. Martone, A. Mons, A. L. Packer, B. Persson, P. Rocca-Serra, M. Roos, R. van Schaik, S.-A. Sansone, E. Schultes, T. Sengstag, T. Slater, G. Strawn, M. A. Swertz, M. Thompson, J. van der Lei, E. van Mulligen, J. Velterop, A. Waagmeester, P. Wittenburg, K. Wolstencroft, J. Zhao, and B. Mons, Scientific Data **3,** 160018 (2016).

[122] B. Mons, C. Neylon, J. Velterop, M. Dumontier, L. O. B. da Silva Santos, and M. D. Wilkinson, Information Services & Use **37,** 49 (2017).

[123] H. Wang, X. D. Xiang, and L. Zhang, Engineering **6,** 609 (2020).

[124] A. Jain, S. P. Ong, G. Hautier, W. Chen, W. D. Richards, S. Dacek, S. Cholia, D. Gunter, D. Skinner, G. Ceder, and K. A. Persson, APL Materials **1,** 011002 (2013).

[125] Atomly, https://atomly.net.

[126] A. Belsky, M. Hellenbrandt, V. L. Karen, and P. Luksch, Acta Crystallographica Section B: Structural Science **58,** 364 (2002).

[127] S. Curtarolo, W. Setyawan, G. L. W. Hart, M. Jahnatek, R. V. Chepulskii, R. H. Taylor, S. Wang, J. Xue, K. Yang, O. Levy, M. J. Mehl, H. T. Stokes, D. O. Demchenko, and D. Morgan, Computational Materials Science **58,** 218 (2012).

[128] P. Hodge, K.-H. Hellwich, R. C. Hiorns, R. G. Jones, J. Kahovec, C. K. Luscombe, M. D. Purbrick, and E. S. Wilks,   **92,** 797 (2020).

[129] C. Lin, P.-H. Wang, Y. Hsiao, Y.-T. Chan, A. C. Engler, J. W. Pitera, D. P. Sanders, J. Cheng, and Y. J. Tseng, ACS Applied Polymer Materials **2,** 3107 (2020).

[130] S. Otsuka, I. Kuwajima, J. Hosoya, Y. Xu, and M. Yamazaki, in *PoLyInfo: Polymer Database for Polymeric Materials Design*, 2011, p. 22.

[131] Khazana, Materials data and tools from the Ramprasad Group, https://khazana.gatech.edu/.

[132] B. Ellis and R. Smith, *Polymers: a property database* (CRC press, 2008).

[133] Polymer Property Predictor and Database, https://pppdb.uchicago.edu/.

[134] CAMPUS - a material information system for the plastics industry, https://www.campusplastics.com/.

[135] Y. Hayashi, J. Shiomi, J. Morikawa, and R. Yoshida, npj Computational Materials **8,** 222 (2022).

[136] C. Kim, A. Chandrasekaran, T. D. Huan, D. Das, and R. Ramprasad, The Journal of Physical Chemistry C **122,** 17575 (2018).

[137] eMolecules, https://www.emolecules.com.

[138] E. W. C. Spotte-Smith, O. A. Cohen, S. M. Blau, J. M. Munro, R. Yang, R. D. Guha, H. D. Patel, S. Vijay, P. Huck, R. Kingsbury, M. K. Horton, and K. A. Persson, Digital Discovery (2023).

[139] M. Meuwly, Chemical Reviews **121,** 10218 (2021).

[140] P. Shetty and R. Ramprasad, Journal of Chemical Information and Modeling **61,** 5377 (2021).

[141] M. Ohno, Y. Hayashi, Q. Zhang, Y. Kaneko, and R. Yoshida, Journal of Chemical Information and Modeling **63,** 5539 (2023).

[142] G. Landrum, in *https://www.rdkit.org/* (2020).

[143] R. L. C. Akkermans, N. A. Spenley, and S. H. Robertson, Molecular Simulation **39,** 1153 (2013).

[144] S. Jo, T. Kim, V. G. Iyer, and W. Im, Journal of Computational Chemistry **29,** 1859 (2008).

[145] L. Martínez, R. Andrade, E. G. Birgin, and J. M. Martínez, Journal of Computational Chemistry **30,** 2157 (2009).

[146] A. I. Jewett, D. Stelter, J. Lambert, S. M. Saladi, O. M. Roscioni, M. Ricci, L. Autin, M. Maritan, S. M. Bashusqeh, T. Keyes, R. T. Dame, J.-E. Shea, G. J. Jensen, and D. S. Goodsell, Journal of Molecular Biology **433,** 166841 (2021).

[147] M. E. Fortunato and C. M. Colina, SoftwareX **6,** 7 (2017).

[148] H. Sahu, K.-H. Shen, J. H. Montoya, H. Tran, and R. Ramprasad, Journal of Chemical Theory and Computation **18,** 2737 (2022).

[149] P. J. in 't Veld and G. C. Rutledge, Macromolecules **36,** 7358 (2003).

[150] A. P. Thompson, H. M. Aktulga, R. Berger, D. S. Bolintineanu, W. M. Brown, P. S. Crozier, P. J. in 't Veld, A. Kohlmeyer, S. G. Moore, T. D. Nguyen, R. Shan, M. J. Stevens, J. Tranchida, C. Trott, and S. J. Plimpton, Computer Physics Communications **271,** 108171 (2022).

[151] Polymer Molecular Dynamics (PMD). https://polymer-molecular-dynamics.netlify.app/.

[152] D. Weininger, Journal of Chemical Information and Computer Sciences **28,** 31 (1988).

[153] M. E. Fisher, The Journal of Chemical Physics **44,** 616 (2004).

[154] D. Chowdhury and B. K. Chakrabarti, Journal of Physics A: Mathematical and General **18,** L377 (1985).

[155] D. Surblys, H. Matsubara, G. Kikugawa, and T. Ohara, Physical Review E **99,** 051301 (2019).

[156] P. Boone, H. Babaei, and C. E. Wilmer, Journal of Chemical Theory and Computation **15,** 5579 (2019).

[157] J. Wang, R. M. Wolf, J. W. Caldwell, P. A. Kollman, and D. A. Case, Journal of Computational Chemistry **25,** 1157 (2004).

[158] X. He, V. H. Man, W. Yang, T. S. Lee, and J. Wang, Journal of Chemical Physics **153** (2020).

[159] X. He, B. Walker, V. H. Man, P. Ren, and J. Wang, Current Opinion in Structural Biology **72,** 187 (2022).

[160] G. S. Larsen, P. Lin, K. E. Hart, and C. M. Colina, Macromolecules **44,** 6944 (2011).

[161] X. Yang, Q. Liu, X. Zhang, C. Ji, and B. Cao, Fluid Phase Equilibria **562,** 113566 (2022).

[162] P. K. Schelling, S. R. Phillpot, and P. Keblinski, Physical Review B **65,** 144306 (2002).

[163] Z. Fan, H. Dong, A. Harju, and T. Ala-Nissila, Physical Review B **99,** 064308 (2019).

[164] F. DeAngelis, M. G. Muraleedharan, J. Moon, H. R. Seyf, A. J. Minnich, A. J. H. McGaughey, and A. Henry, Nanoscale and Microscale Thermophysical Engineering **23,** 81 (2019).

[165] J. E. Turney, E. S. Landry, A. J. H. McGaughey, and C. H. Amon, Physical Review B **79,** 064301 (2009).

[166] F. Müller-Plathe, The Journal of Chemical Physics **106,** 6082 (1997).

[167] D. Torii, T. Nakano, and T. Ohara, The Journal of Chemical Physics **128,** 044504 (2008).

[168] Y. Zhao, R. J. Mulder, S. Houshyar, and T. C. Le, Polymer Chemistry **14,** 3325 (2023).

[169] S. Stuart, J. Watchorn, and F. X. Gu, npj Computational Materials **9,** 102 (2023).

[170] D. Weininger, A. Weininger, and J. L. Weininger, Journal of Chemical Information and Computer Sciences **29,** 97 (1989).

[171] M. Krenn, F. Häse, A. Nigam, P. Friederich, and A. Aspuru-Guzik, Machine Learning: Science and Technology **1,** 045024 (2020).

[172] A. Yüksel, E. Ulusoy, A. Ünlü, and T. Doğan, Machine Learning: Science and Technology **4,** 025035 (2023).

[173] L. Turcani, E. Berardo, and K. E. Jelfs, Journal of Computational Chemistry **39,** 1931 (2018).

[174] M. M. Cencer, J. S. Moore, and R. S. Assary, Polymer International **71,** 537 (2022).

[175] R.-R. Griffiths and J. M. Hernández-Lobato, Chemical Science **11,** 577 (2020).

[176] C. Kuenneth and R. Ramprasad, Nature Communications **14,** 4099 (2023).

[177] C. Xu, Y. Wang, and A. Barati Farimani, npj Computational Materials **9,** 64 (2023).

[178] H. Qiu, L. Liu, X. Qiu, X. Dai, X. Ji, and Z.-Y. Sun, Chemical Science (2024).

[179] J. L. Durant, B. A. Leland, D. R. Henry, and J. G. Nourse, Journal of Chemical Information and Computer Sciences **42,** 1273 (2002).

[180] E. L. Willighagen, J. W. Mayfield, J. Alvarsson, A. Berg, L. Carlsson, N. Jeliazkova, S. Kuhn, T. Pluskal, M. Rojas-Chertó, O. Spjuth, G. Torrance, C. T. Evelo, R. Guha, and C. Steinbeck, Journal of Cheminformatics **9,** 33 (2017).

[181] N. M. O'Boyle, M. Banck, C. A. James, C. Morley, T. Vandermeersch, and G. R. Hutchison, Journal of Cheminformatics **3,** 33 (2011).

[182] X. Zhang, G. Wei, Y. Sheng, W. Bai, J. Yang, W. Zhang, and C. Ye, ACS Applied Materials & Interfaces **15,** 21537 (2023).

[183] A. Capecchi, D. Probst, and J.-L. Reymond, Journal of Cheminformatics **12,** 43 (2020).

[184] H. L. Morgan, Journal of chemical documentation **5,** 107 (1965).

[185] D. Rogers and M. Hahn, Journal of Chemical Information and Modeling **50,** 742 (2010).

[186] S. Jaeger, S. Fulle, and S. Turk, Journal of Chemical Information and Modeling **58,** 27 (2018).

[187] R. Ma, Z. Liu, Q. Zhang, Z. Liu, and T. Luo, Journal of Chemical Information and Modeling **59,** 3110 (2019).

[188] L. Tao, J. Byrnes, V. Varshney, and Y. Li, iScience **25,** 104585 (2022).

[189] M. Karelson, V. S. Lobanov, and A. R. Katritzky, Chemical Reviews **96,** 1027 (1996).

[190] A. Mauri, V. Consonni, M. Pavan, and R. Todeschini, Match **56,** 237 (2006).

[191] M. Haghighatlari, G. Vishwakarma, D. Altarawy, R. Subramanian, B. U. Kota, A. Sonpal, S. Setlur, and J. Hachmann, WIREs Computational Molecular Science **10,** e1458 (2020).

[192] H. Moriwaki, Y.-S. Tian, N. Kawashita, and T. Takagi, Journal of Cheminformatics **10,** 4 (2018).

[193] J. S. Gábor, L. R. Maria, and K. B. Nail, The Annals of Statistics **35,** 2769 (2007).

[194] D. Albanese, M. Filosi, R. Visintainer, S. Riccadonna, G. Jurman, and C. Furlanello, Bioinformatics **29,** 407 (2013).

[195] I. Guyon, J. Weston, S. Barnhill, and V. Vapnik, Machine Learning **46,** 389 (2002).

[196] E. Bisong, in *Building Machine Learning and Deep Learning Models on Google Cloud Platform:*

*A Comprehensive Guide for Beginners*, edited by E. Bisong (Apress, Berkeley, CA, 2019), p. 287.

[197] S. M. Lundberg and S.-I. Lee, Advances in neural information processing systems **30,** 4765−4774 (2017).

[198] C. Loftis, K. Yuan, Y. Zhao, M. Hu, and J. Hu, The Journal of Physical Chemistry A **125,** 435 (2021).

[199] T. Stephens, in *https://gplearn.readthedocs.io/en/stable/*.

[200] Y. Wang, N. Wagner, and J. M. Rondinelli, MRS Communications **9,** 793 (2019).

[201] T. A. R. Purcell, M. Scheffler, L. M. Ghiringhelli, and C. Carbogno, npj Computational Materials **9,** 112 (2023).

[202] M. Cranmer, arXiv preprint arXiv:2305.01582 (2023).

[203] R. Ouyang, E. Ahmetcik, C. Carbogno, M. Scheffler, and L. M. Ghiringhelli, Journal of Physics: Materials **2,** 024002 (2019).

[204] Examples of symbolic regressor in gplearn. https://gplearn.readthedocs.io/en/stable/examples.html.

[205] pysr_demo. https://github.com/MilesCranmer/PySR/blob/master/examples/pysr_demo.ipynb.

[206] Tutorial for SISSO. https://gitlab.mpcdf.mpg.de/nomad-lab/ai-toolkit/tutorial-compressed-sensing/-/blob/master/compressed_sensing.ipynb.

[207] F. Pedregosa, G. Varoquaux, A. Gramfort, V. Michel, B. Thirion, O. Grisel, M. Blondel, P. Prettenhofer, R. Weiss, and V. Dubourg, the Journal of machine Learning research **12,** 2825 (2011).

[208] F. Nogueira, in *https://github.com/fmfn/BayesianOptimization* (2014).

[209] A. Tiihonen, S. J. Cox-Vazquez, Q. Liang, M. Ragab, Z. Ren, N. T. P. Hartono, Z. Liu, S. Sun, C. Zhou, N. C. Incandela, J. Limwongyut, A. S. Moreland, S. Jayavelu, G. C. Bazan, and T. Buonassisi, Journal of the American Chemical Society **143,** 18917 (2021).

[210] F. R. Burden, Quantitative Structure-Activity Relationships **16,** 309 (1997).

[211] P. Mohamad Zaim Awang and K. K. Krishna Prakash, Sparklinglight Transactions on Artificial Intelligence and Quantum Computing (STAIQC) **1,** 36 (2021).

[212] M. Abadi, A. Agarwal, P. Barham, E. Brevdo, Z. Chen, C. Citro, G. S. Corrado, A. Davis, J. Dean, and M. Devin, arXiv preprint arXiv:1603.04467 (2016).

[213] X. Huang, C. Y. Zhao, H. Wang, and S. Ju, arXiv preprint arXiv:2402.11600 (2024).

[214] B. Weng, Z. Song, R. Zhu, Q. Yan, Q. Sun, C. G. Grice, Y. Yan, and W.-J. Yin, Nature

Communications **11,** 3513 (2020).

[215] K. Y. L. Andre, V.-G. Eleonore, L. Yee-Fun, and H. Kedar, Journal of Materials Informatics **3,** 11 (2023).

[216] G. Agarwal, H. A. Doan, L. A. Robertson, L. Zhang, and R. S. Assary, Chemistry of Materials **33,** 8133 (2021).

[217] P. Ertl and A. Schuffenhauer, Journal of Cheminformatics **1,** 8 (2009).

[218] M. T. Bhoskar, M. O. K. Kulkarni, M. N. K. Kulkarni, M. S. L. Patekar, G. M. Kakandikar, and V. M. Nandedkar, Materials Today: Proceedings **2,** 2624 (2015).

[219] X. Huang, S. Ma, H. Wang, S. Lin, C. Y. Zhao, H. Wang, and S. Ju, International Journal of Heat and Mass Transfer **197,** 123332 (2022).

[220] J. Blank and K. Deb, IEEE Access **8,** 89497 (2020).

[221] K. Deb, K. Sindhya, and T. Okabe, in *Proceedings of the 9th annual conference on Genetic and evolutionary computation* (Association for Computing Machinery, London, England, 2007), p. 1187.

[222] T. Ueno, T. D. Rhone, Z. Hou, T. Mizoguchi, and K. Tsuda, Materials discovery **4,** 18 (2016).

[223] S. Ju, S. Shimizu, and J. Shiomi, Journal of Applied Physics **128,** 161102 (2020).

[224] A. Agnihotri and N. Batra, Distill **5,** e26 (2020).

[225] M. Balandat, B. Karrer, D. Jiang, S. Daulton, B. Letham, A. G. Wilson, and E. Bakshy, Advances in neural information processing systems **33,** 21524 (2020).

[226] P. Ray, B. K. Chakrabarti, and A. Chakrabarti, Physical Review B **39,** 11828 (1989).

[227] T. Kadowaki and H. Nishimori, Physical Review E **58,** 5355 (1998).

[228] Z. Mao, Y. Matsuda, R. Tamura, and K. Tsuda, Digital Discovery **2,** 1098 (2023).

[229] B. A. Wilson, Z. A. Kudyshev, A. V. Kildishev, S. Kais, V. M. Shalaev, and A. Boltasseva, Applied Physics Reviews **8,** 041418 (2021).

[230] K. Kitai, J. Guo, S. Ju, S. Tanaka, K. Tsuda, J. Shiomi, and R. Tamura, Physical Review Research **2,** 013319 (2020).

[231] dimod. https://github.com/dwavesystems/dimod.

[232] K. Deb and H. Jain, IEEE Transactions on Evolutionary Computation **18,** 577 (2014).

[233] H. Jain and K. Deb, IEEE Transactions on Evolutionary Computation **18,** 602 (2014).

[234] M. Seifrid, R. Pollice, A. Aguilar-Granda, Z. Morgan Chan, K. Hotta, C. T. Ser, J. Vestfrid, T. C. Wu, and A. Aspuru-Guzik, Accounts of Chemical Research **55,** 2454 (2022).

[235] A. A. Volk, R. W. Epps, D. T. Yonemoto, B. S. Masters, F. N. Castellano, K. G. Reyes, and M. Abolhasani, Nature Communications **14,** 1403 (2023).

[236] D. A. Boiko, R. MacKnight, B. Kline, and G. Gomes, Nature **624,** 570 (2023).